\documentclass[%
reprint,
twocolumn,
amsmath,amssymb,
aip,pop,
nofootinbib,
longbibliography, 
floatfix,
]{revtex4-2}
\usepackage[utf8]{inputenc}
\usepackage{lipsum}
\usepackage{graphicx}
\usepackage{dcolumn}
\usepackage{bm}
\usepackage{amssymb}
\usepackage{amsmath}
\usepackage{textcomp}
\usepackage{color}
\usepackage{mathtools}
\usepackage{units}
\usepackage{appendix}

\begin{document}

\title{Particle-in-cell simulations of expanding high energy density plasmas with laser ray tracing}

\author{K. V. Lezhnin}
\email{klezhnin@pppl.gov}
\affiliation{Princeton Plasma Physics Laboratory, 100 Stellarator Rd, Princeton, NJ 08540, USA}
\author{S. R. Totorica}
\affiliation{Department of Astrophysical Sciences, Princeton University, Princeton, NJ 08544, USA}
\affiliation{Department of Astro-fusion Plasma Physics (AFP), Headquarters for Co-Creation Strategy, National Institutes of Natural Sciences, Tokyo 105-0001, Japan}
\author{A. S. Hyder}
\affiliation{Department of Applied Physics and Applied Mathematics, Columbia University, New York, NY 10027, USA}
\author{J. Griff-McMahon}
\affiliation{Princeton Plasma Physics Laboratory, 100 Stellarator Rd, Princeton, NJ 08540, USA}
\affiliation{Department of Astrophysical Sciences, Princeton University, Princeton, NJ 08544, USA}
\author{M. B. P. Adams}
\affiliation{Sandia National Laboratories, Albuquerque, NM 87185, USA}
\author{P. Tzeferacos}
\affiliation{Department of Physics and Astronomy, University of Rochester, Rochester, NY 14627, USA}
\author{A. Diallo}
\affiliation{Princeton Plasma Physics Laboratory, 100 Stellarator Rd, Princeton, NJ 08540, USA}
\author{W. Fox}
\affiliation{Princeton Plasma Physics Laboratory, 100 Stellarator Rd, Princeton, NJ 08540, USA}
\affiliation{Department of Astrophysical Sciences, Princeton University, Princeton, NJ 08544, USA}

\date{\today}

\begin{abstract}
The design and analysis of high energy density (HED) laser experiments typically rely on radiation hydrodynamics simulations. However, some laser-plasma interaction regimes are not collisional and cannot be adequately modeled with hydrodynamics. For example, strongly driven magnetic reconnection and magnetized collisionless shock experiments possess extended hydrodynamic or even kinetic properties, necessitating first-principles kinetic simulations. In this paper, we present the benchmarking and first results obtained with a laser-ray-tracing and inverse Bremsstrahlung absorption module implemented in the particle-in-cell code PSC. The simulation results are compared to radiation hydrodynamic simulations using the FLASH code as well as analytical estimates. We successfully benchmark the energy deposition model and overall hydrodynamic evolution of the systems. We also consider possible kinetic effects that may be expected from laser-target ablation in the HED regime, including non-local transport and two-temperature effects.

\end{abstract}

\maketitle

\section{Introduction}\label{sec:intro}

High energy density (HED) physics is an actively developing field of plasma physics with a multitude of applications, including inertial fusion energy (IFE) \cite{Betti2016}, stockpile stewardship \cite{Reis2016}, and laboratory astrophysics \cite{Takabe2021}. Recent successes at the National Ignition Facility of achieving scientific gain larger than one has marked a significant step towards IFE \cite{Zylstra2022}. However, maintaining rapid progress in HED science is challenging, since large laser facilities required for HED physics experiments are scarce, diagnostics are sophisticated and often limited by the experimental design, and the shot repetition rate of large laser facilities is on the order of $10^1$ per day at best. This is where numerical simulations show their value, providing a complementary approach to the design and interpretation of HED experiments. Developing high-fidelity simulation codes is a difficult task and requires contributions from software engineering, high performance computing, and physics experts. The codes must be verified, validated, and benchmarked against other codes and experimental data \cite{Orban2021,Orban2022}.

The current state-of-the-art numerical tool in HED science is radiation hydrodynamics simulation. Radiation hydrodynamics simulations played a key role in optimizing the laser and target parameters for the recent success at the National Ignition Facility\cite{HYDRAsims}. It models the plasma as a compressible fluid, and typically includes additional physics such as thermal transport, radiation transport, and laser absorption. Certain approaches can support magnetized plasma effects, including Biermann battery magnetic field generation \cite{Pilgram2022} and the Nernst effect \cite{Walsh2020}. The system of hydrodynamic equations is closed by introducing an equation of state (EOS) to calculate the thermodynamic potentials and calculating the mean ionization as a function of density and temperature, which are usually tabulated using EOS/opacity codes like IONMIX \cite{Macfarlane1989} or PrOpacEOS \cite{PrOpacEOS}. The aforementioned approach is implemented in various codes, including FLASH \cite{Fryxell2000,Tzeferacos2015}, RALEF\cite{RALEF}, HYDRA\cite{HYDRA}, DRACO\cite{DRACO}, STAR\cite{STAR}, and HELIOS\cite{HELIOS}.

Despite the successful history of hydrodynamic modeling in HED science, there is strong evidence that HED plasmas possess extended hydrodynamic or even kinetic properties (see, e.g., Ref.~\onlinecite{Rinderknecht2018}). The hydrodynamic assumption of high collisionality and classical thermal transport is violated for sufficiently hot and rarified plasmas, commonly found in HED environments. Indeed, long pulses at large laser facilities typically heat electrons to the keV temperature range \cite{Manheimer1982} and deposit energy at the critical electron density ($n_e=n_{\rm cr} = 1.1\cdot 10^{21}[\lambda/1 \rm \mu m]^{-2}\,\rm cm^{-3}$). The resulting electron mean free path, $\lambda_{\rm e,mfp}$, is on the order of $10\, \mu$m (for hydrogen plasma) and is comparable to typical kinetic length scales such as the ion inertial scale $d_{i}=c/\omega_{pi}$. Furthermore, large $\lambda_{\rm e,mfp}$ values infringe on the basic assumptions of hydrodynamics (strongly collisional plasma, $\lambda_{\rm mfp}/L \ll 1$, where $L$ is the size of the system) and classical thermal transport ($\lambda_{\rm mfp}/L_T \ll 1$, where $L_T \equiv T_e / |dT_e/dx|$ is the local electron temperature gradient length scale). For this reason, extended hydrodynamics approaches with nonlocal transport models\cite{SNB}, hybrid approaches \cite{HybridVPIC,LSP} (e.g., kinetic treatment of ions and fluid treatment of electrons), or first-principles kinetic approaches \cite{Oshun} have been proposed to more accurately capture the physics of such plasma conditions.

Some of the extended hydrodynamics approaches were recently implemented in hydrodynamic codes, showing success in capturing the physics by utilizing closures obtained from first-principles calculations, see Ref.~\cite{Rinderknecht2018} and references therein for examples from the ICF field. On the other hand, recent work by Ridgers et al.\cite{Ridgers2020} clearly outlined the discrepancy in Biermann battery physics between extended magnetohydrodynamic (MHD) and kinetic modeling. Experimental measurements of Biermann battery magnetic fields also highlighted deviations from extended MHD simulations\cite{GriffMcMahon2023}. Kinetic modeling is important to predict Biermann magnetic field generation and to benchmark extended MHD models.

The study of non-Maxwellian features is another class of HED experiments that requires kinetic physics. This includes laboratory studies of the Weibel instability\cite{Fox2013}, particle acceleration by magnetic reconnection \cite{Totorica2016,Totorica2017,Fox2017,Totorica2020,Chien2023}, and magnetized \cite{Schaeffer2017} or Weibel-mediated\cite{Fiuza2020} collisionless shocks. Strong non-Maxwellian populations can alter or even dominate transport properties in {a} plasma\cite{Bell1981,Matte1982,Mora1982}. Plasma interpenetration and anisotropy-driven instabilities represent another category of non-Maxwellian effects that is typically not captured in extended hydrodynamic modeling, but can play a large role in HED plasma dynamics. For instance, in Fiuza et al.\cite{Fiuza2020}, plasma interpenetration and the development of the Weibel instability was the primary mechanism for the formation of the collisionless shock and for electron acceleration in the shock downstream. Similarly, in Fox et al.\cite{Fox2021}, anisotropy-driven instabilities modulated the current sheet in the magnetic reconnection experiment and led to the onset of a novel reconnection regime. Finally, laser absorption through the inverse Bremsstrahlung (IB) mechanism is known to generate non-Maxwellian features and modify the laser absorption efficiency through the Langdon effect\cite{Langdon1980}. Recent experimental work by Turnbull et al.\cite{Turnbull2023} highlighted the importance of kinetic corrections to the classical IB to match the experimentally measured absorption.

Kinetic simulations have previously been used in the HED field to achieve accurate physics results and obtain closures for extended hydrodynamics and hybrid models\cite{Rinderknecht2018}. The two primary approaches to solving fully kinetic physics are Vlasov Fokker-Planck (VFP) and Particle-In-Cell (PIC) methods. The latter is better suited for simulating laser-plasma interactions with large temperature gradients and multiple spatial dimensions\cite{Thomas2012}, as is common in HED experiments. To capture HED-related physics in PIC simulations, the model should include the dynamics of laser heating through laser ray tracing, dynamic ionization, and radiation transport. These modules have been implemented in various PIC codes in the HED field\cite{Rinderknecht2018} as well as short pulse physics\cite{EPOCH,SMILEI} and X-ray laser simulations\cite{Sentoku2014}. Prior PIC simulations have considered laser-plasma interactions by resolving laser wavelength with the numerical grid (see, e.g., Ref.~\onlinecite{EPOCH}); however this full-wave modeling limited the temporal timescales of the simulation. Here we consider a laser ray-tracing approach, which makes simulating temporal and spatial scales of HED experiments feasible. This work extends a previously-utilized volumetric heating operator within PSC \cite{Fox2018}, which does not track the critical surface location and may misalign the spatial laser energy deposition, possibly generating unphysical expanding plasma profiles.

In this work, we use the particle-in-cell code PSC\cite{Germaschewski2016} in conjunction with a recently developed laser absorption module\cite{Hyder2024} to study the laser ablation of aluminum in the HED regime, with laser intensity of $\sim 10^{13} \, \rm W/cm^2$. We focus our comparisons and benchmarking on the lower-density, high-temperature ablated plasma region, rather than the compressed solid-density matter which is also produced in laser-solid interaction. The results are compared to FLASH simulations and the steady-state ablation model\cite{Manheimer1982}, finding good overall agreement. We also highlight differences between the models, particularly kinetic effects that might arise during the laser pulse duration. This includes the violation of classical thermal transport assumptions, contributions to generalized Ohm's law, and non-Maxwellian features in the electron distribution.

The paper is organized as follows. In Section~\ref{sec:setup}, we describe the simulation setup used to model aluminum ablation with FLASH and PSC. Section~\ref{sec:results} is devoted to a comparison between FLASH, PSC, and a steady-state ablation model as well as an estimation of the role of kinetic effects. In Section~\ref{sec:discussion}, we explore the limitations of the present PSC model and discuss the applications of PIC with laser ray tracing, including Biermann battery field generation and extreme ultraviolet (EUV) source simulations\cite{Totorica2024}. Lastly, we address future steps and possible extensions to the current PSC model, as well as document our convergence test studies with respect to reduced simulation parameters and parameters of the Coulomb collision module.

\section{Simulation setup}\label{sec:setup}

\begin{figure*}
    \centering
    \includegraphics[width=\linewidth]{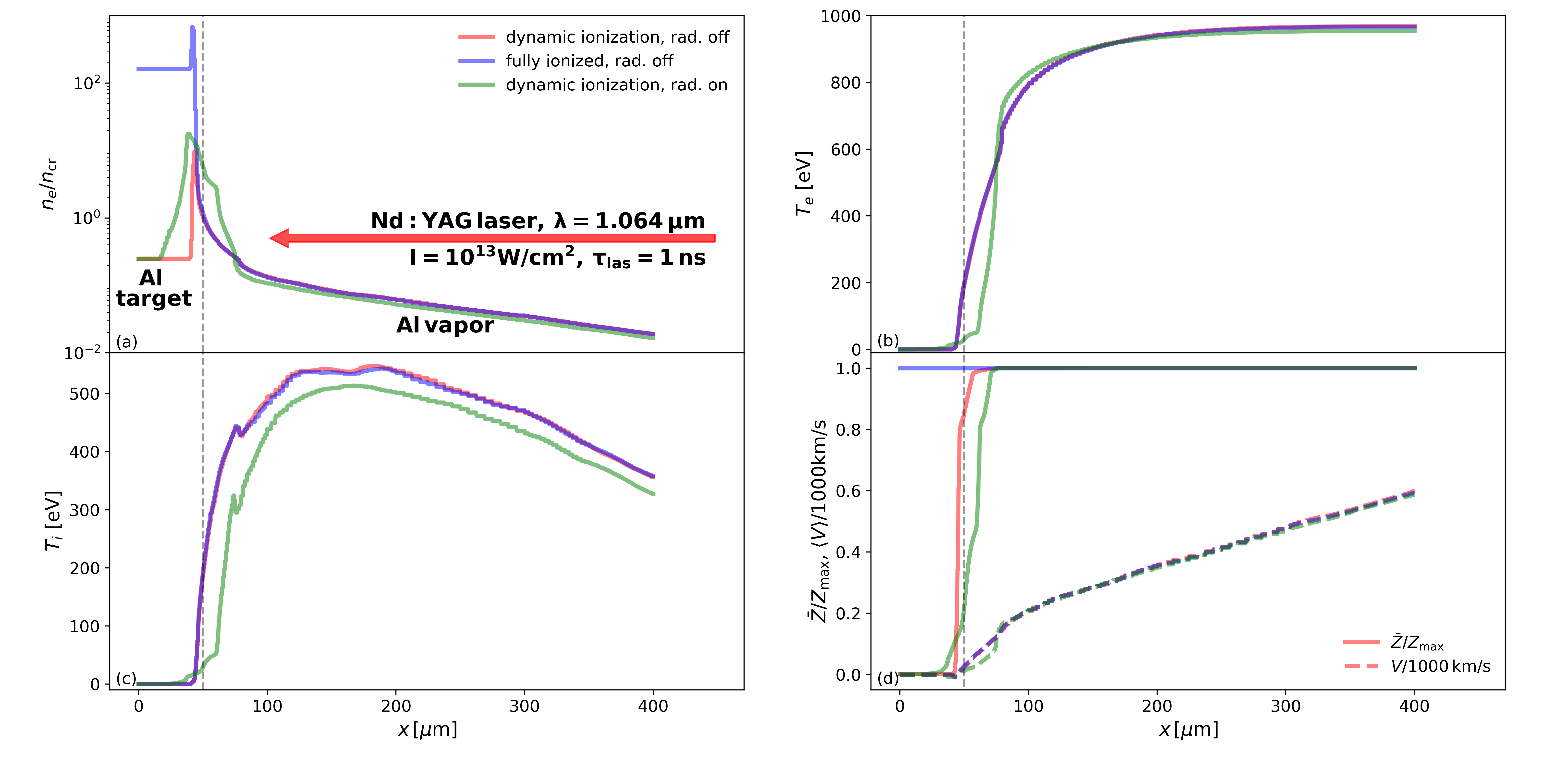}
    \caption{Simulation setup: $\lambda = 1.064\, \rm \mu m$, $10^{13}\, \rm W/cm^2$, $1\, \rm ns$ laser pulse interacts with solid Al target. (a) Electron density profiles at $t=1\rm \, ns$ in FLASH simulations under different ionization and radiation configurations: with dynamic ionization but without radiation transport (red), without ionization or radiation (blue), with ionization and radiation transport (green). The dashed vertical, gray line corresponds to the target edge at $t=0$. (b) Electron temperature with the same color legend. (c) Ion temperature. (d) Ionization fraction (solid line) and ion velocity (dashed line). Radiation transport and dynamic ionization have a minimal impact on plasma profiles.}
    \label{fig:flash_init}
\end{figure*}

Let us start by describing the simulation setup implemented in the FLASH and PSC simulations. We consider a one-dimensional, solid aluminum target, a few tens of microns thick, irradiated with a long-pulse laser. This generates a hot and tenuous expanding plasma, see Figure~\ref{fig:flash_init}. To achieve an HED plasma regime (plasma pressure $\sim 1\,\rm Mbar$), we use typical parameters for a high-energy, long-pulse optical laser. The laser has an intensity of $10^{13}\, \rm W/cm^2$, wavelength of $1.064\, \mu \rm m$, and pulse duration $\sim1$ ns. We assume a spatially uniform intensity, and the laser is normally incident on the target. Since the current version of PSC lacks some physics implemented in FLASH (e.g., ionization and radiation transport), we also compare FLASH with and without these effects to quantify the discrepancy between more realistic and idealized plasma behavior.

\subsection{FLASH}

In FLASH, we start with the ``LaserSlab" example problem within the code, which uses single-fluid two-temperature model in conjunction with the equations of state for both electron and ion components of the fluid. The problem also includes additional physics modules such as electron thermal transport, collisional electron-ion equilibration, radiation transport, laser ray-tracing, and laser energy deposition. The system of equations is closed using tabulated equations of state. The FLASH simulations in this work disable radiation transport and assume a fully ionized aluminum target at t=0 in order to directly compare with PSC, which currently lacks these features. Three FLASH simulations were conducted with different radiation and ionization configurations to determine their effects on the ablated plasma behavior. Figure~\ref{fig:flash_init} shows the spatial profiles of electron density, electron temperature, ion temperature, ionization fraction, and ion velocity (a-d) from these runs at t=1 ns. The red curves included ionization but disabled radiation transport, the blue curve disabled both effects, and the green curve included both effects. The negligible differences between the simulations affirm that excluding ionization and radiation effects in PIC is reasonable and leads to accurate predictions of plasma ablation in the simplified FLASH model.

For reference and to clarify the comparison with PSC, we provide the system of equations that are solved by FLASH\cite{Tzeferacos2015}:

\begin{align}
    &\frac{\partial}{\partial t}\rho + \nabla \cdot (\rho \vec{v}) = 0, \label{eqn:cont} \\
    &\frac{\partial}{\partial t}\vec{v} + (\vec{v}\cdot \nabla)\vec{v} = - \frac{1}{\rho} \nabla (P_{e}+P_{i}), \label{eqn:momentum} \\
    &\frac{\partial}{\partial t}e_i+(\vec{v}\cdot \nabla)e_i = -\frac{P_i}{\rho}\nabla \cdot \vec{v}+ \frac{C_{V,e}}{\tau_{ei}^{\mathcal{E}}} (T_e-T_i), \label{eqn:eion} \\
    &\frac{\partial}{\partial t}e_e+(\vec{v}\cdot \nabla)e_e = -\frac{P_e}{\rho}\nabla \cdot \vec{v}+ \frac{C_{V,e}}{\tau_{ei}^{\mathcal{E}}} (T_i-T_e) \nonumber \\
    &-\frac{1}{\rho} \nabla \cdot \vec{q_e} + \frac{Q_{\rm las}}{\rho}, \label{eqn:eele} \\
    &\vec{q_e} = - {\rm sign(\nabla T_e)} ((K_{\rm ele} \nabla T_e)^{-2}+q_{\rm max}^{-2} )^{-1/2}. \label{eqn:heatflux}
\end{align}

\noindent Here, $\rho$ is the plasma mass density, $\vec{v}$ is the flow speed, $P_e$ and $P_i$ are the electron and ion pressures, $e_i$ and $e_e$ are the specific internal energy of ions and electrons, $C_{V,e}$ is the isochoric heat capacity of electrons, $T_e$ and $T_i$ are the electron and ion temperatures, $\tau_{ei}^{\mathcal{E}}$ is the electron-ion collisional temperature equilibration time, $q_e$ is the electron heat flux, $K_e$ is the electron heat conductivity, $q_{\rm max}$ is the limiting heat flux, and $Q_{\rm las}$ is the energy source due to laser heating. Note that the heat flux $\vec{q_e}$ is calculated using the Larsen flux limiter. The system of Eqs.~\ref{eqn:cont}-\ref{eqn:heatflux} is closed with analytical expressions for $K_{\rm ele}$ (Spitzer heat conductivity), $q_{\rm max}$ (limiting heat flux), $\tau_{ei}^{\mathcal{E}}$ (electron-ion collisional heat exchange), and $Q_{\rm las}$ (inverse Bremsstrahlung absorption)\cite{NRLFormulary}. Additionally, thermodynamic potentials (pressure, heat capacity, specific internal energy) and mean ionization $\bar{z}$ are calculated via the tabulated EOS for given values of $\rho$ and $T_{e,i}$. Below, we detail the expressions for $K_e$, $q_{\rm max}$, $\tau_{ei}^{\mathcal{E}}$, and $Q_{\rm las}$:

\begin{align}
    & K_{e} =\left(\frac{8}{\pi}\right)^{3/2} \frac{k_B^{7/2}}{e^4\sqrt{m_e}}\left(\frac{1}{\bar{z}+3.3}\frac{T_e^{5/2}}{\ln \Lambda_{ei}} \right), \label{eqn:Kele}\\
    & q_{\rm max} = \alpha_{\rm ele} n_e k_B^{3/2} T_e^{3/2}/m_e^{1/2}, \label{eqn:qmax}\\
    &\tau_{ei}^{\mathcal{E}} = \frac{3 k_B^{3/2} m_e^{1/2}}{8 \sqrt{2 \pi} e^4} \left( \frac{m_i}{m_e} \right)^{-1/2} \frac{(T_e \,m_i/m_e +T_i)^{3/2}}{\bar{z}^2n_i \ln \Lambda_{ei}}, \label{eqn:tauei} \\
    & Q_{\rm las} = \nu_{\rm IB} P_{\rm las}, \label{eqn:Plas} \\
    & \nu_{\rm IB} = \frac{4}{3}\left(\frac{2\pi}{m_e}\right)^{1/2}\frac{n_e \bar{z}e^4 \ln \Lambda_{\rm IB}}{(k_BT_e)^{3/2}}. \label{eqn:nuIB}
\end{align}

\noindent Here, $\alpha_{\rm ele}$ is a electron conductivity flux-limiter coefficient ranging from 0.01 to 0.1. For our simulations, we adopted $\alpha_{\rm ele}=0.06$. $P_{\rm las}$ is the cumulative laser power, $k_B$ is the Boltzmann constant, $e$ is the elementary electron charge, $m_e$ is the electron mass, $m_i$ is the ion mass, and $\nu_{\rm IB}$ is the inverse Bremsstrahlung absorption rate. 

We use nearly the same Coulomb logarithm expression for transport ($\ln \Lambda_{ei}$) and absorption ($\ln \Lambda_{\rm IB}$) theories, which is a simplification that should be ultimately discarded, as suggested by recent experimental measurements of IB\cite{Turnbull2023}. Corrections to this model include (I) using the laser frequency in the calculation of the maximum impact parameter (it should be $b_{\rm max} = v_{\rm th}/\omega = \sqrt{k_BT_e/m_e\omega^2}$ instead of $b_{\rm max} = v_{\rm th}/\omega_{\rm pe} = \sqrt{k_BT_e/4\pi e^2 n_e}$), (II) Langdon and (III) screening effects. As our goal was to compare PSC with FLASH, we used the currently available IB operator in FLASH. However, future detailed benchmarking against experiments should consider these corrections. The exact expressions used in this work are as follows:

\begin{align}
    &\ln \Lambda_{ei} = \ln{\left(\frac{b_{\rm max}}{b_{\rm min}}\right)}, \label{eqn:lnl1} \\
    & b_{\rm max} = \sqrt{\frac{k_B T_e}{4\pi e^2n_e}}, \label{eqn:lnl2} \\
    & b_{\rm min} = {\rm max}\left(\frac{\bar{z}e^2}{3k_BT_e},\frac{\hbar}{2\sqrt{3k_B T_e m_e}} \right), \label{eqn:lnl3}  \\
    &\ln \Lambda_{\rm IB} = \ln \left[\frac{3}{2Ze^2}\frac{k_B^{3/2}T_e^{3/2}}{\pi^{1/2}n_e^{1/2}} \right]. \label{eqn:lnl4}
\end{align}

\noindent $b_{\rm min}$ and $b_{\rm max}$ are the minimum and maximum impact parameters, and $\hbar$ is the reduced Planck constant. These $\ln \Lambda$ expressions are also bounded from below by unity.

The hydrodynamic equations are solved on a rectangular 2D grid using the unsplit hydro solver with third-order reconstruction, the minmod slope limiter, an HLLC Riemann solver, and outflow boundary conditions. Electron thermal transport is solved fully implicitly with the Larsen flux limiter and Neumann boundary conditions. Exponential relaxation electron-ion heat exchange is also used. The simulation grid is cartesian and has a size of $800\, \rm \mu m \times 100\, \rm \mu m$, divided into $8\times 1$ blocks containing $16 \times 16$ cells. The maximum level of adaptive mesh refinement is set to 4. The CFL number is 0.4 and the initial timestep is set at $dt=10^{-15}\, \rm s$ to facilitate the ``slow start" in the simulation\cite{Zhao2023}. The initial conditions set a solid aluminum target with mass density 2.7 $\rm g/cm^3$ from $x=0$ to $50\, \rm \mu m$. The target is fully ionized ($\bar{Z}=13$) and has an initially small temperature of $T_e=T_i=290\, \rm K$. The rest of the box is filled with aluminum vapor at a mass density of $10^{-10} \rm g/cm^3$. The laser pulse is normally incident onto the solid target with a uniform transverse spatial profile, laser wavelength of $\lambda = 1.064\, \rm \mu m$, intensity of $I=10^{13}\, \rm W/cm^2$, and pulse duration of 0.9 ns with the flat-top profile, preceeded by a 0.1 ns linear rise time. We use the tabulated EOS for aluminum, generated using the IONMIX code\cite{Macfarlane1989} and included with FLASH 4.6.2 version.

\subsection{PSC}

PIC simulations are conducted using the PSC code with a recently implemented laser ray tracing module\cite{Hyder2024}. This module models laser ray propagation and energy deposition in the plasma. 

Every few timesteps, we consider 1D laser ray propagation along one of the axes of the numerical grid, and calculate laser absorption according to the IB rate (Eq.~\ref{eqn:Plas},\ref{eqn:nuIB}) up to the critical surface $n_e=n_{\rm cr}$. At this point, the ray is reflected and propagates backwards (i.e., antiparallel to the incident ray), also depositing its energy onto electrons. To model electron heating,  random normally distributed momentum kicks are given to electrons based on the local laser power deposited per particle. This is applied either to a single ray in 1D or to a number of rays corresponding to the transverse grid size in 2D. The validity and limitations of this approach are discussed in Section~\ref{sec:discussion} and further details of the implementation may be found in Ref.~\onlinecite{Hyder2024}.

\begin{figure}
    \centering
    \includegraphics[width=\linewidth]{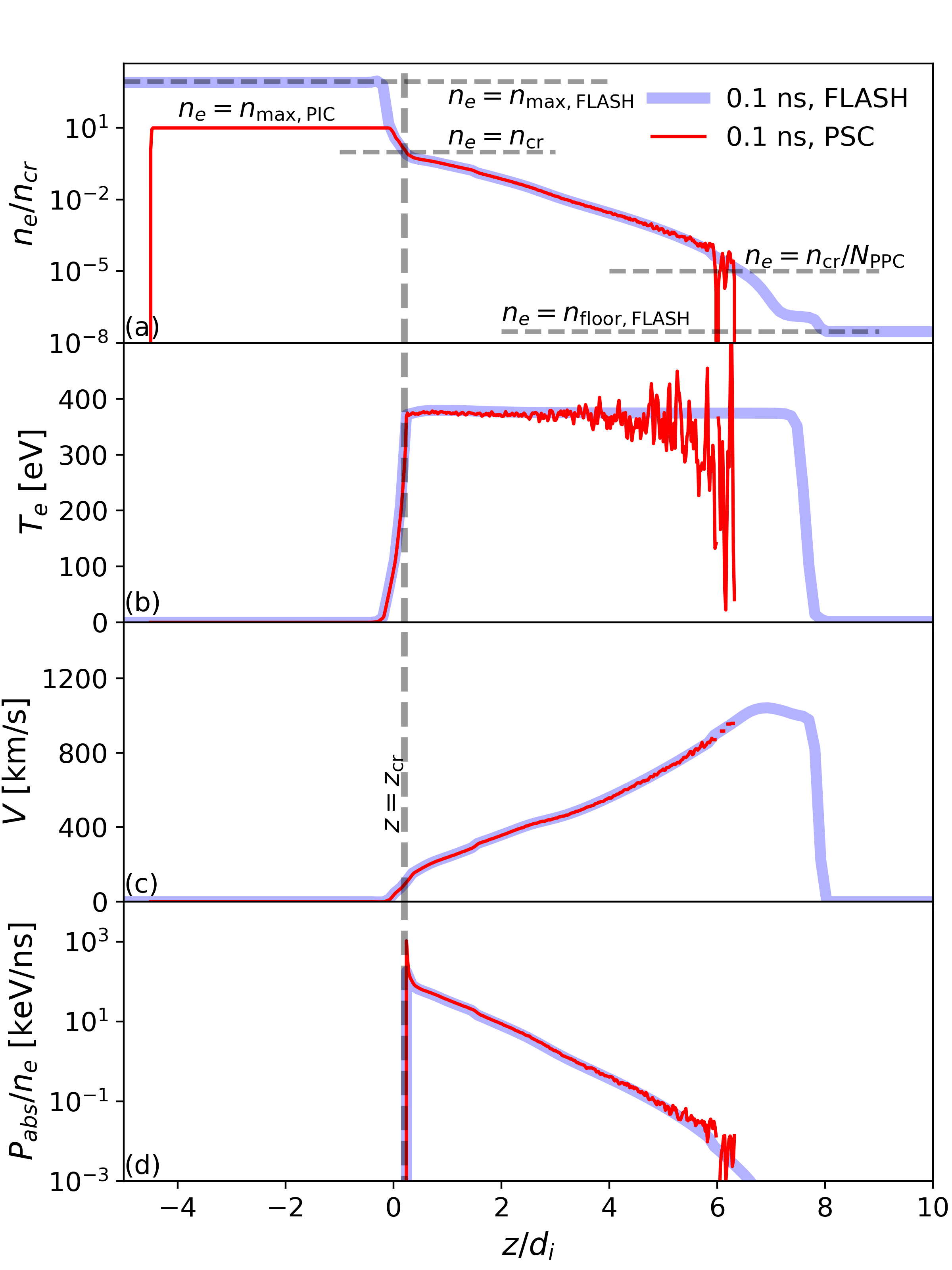}
    \caption{Initial profiles of 1D PSC simulations (red) compared to a FLASH snapshot (blue). Electron density (a), electron temperature (b), and flow speed profiles (c) are matched, with noted differences caused by (I) the capped maximum density in PIC $n_{\rm max,PIC}\ll n_{\rm max,FLASH}$ due to computational constraints and (II) the minimum density in PIC being limited to $n_{\rm cr}/N_{\rm PPC} = 10^{-5} n_{\rm cr} \gg n_{\rm floor,FLASH}$. We also obtain identical laser power absorption profiles (d) in both PSC and FLASH simulations.}
    \label{fig:laserabs}
\end{figure}

We initialize our 1D PIC simulations from the $t=0.1$ ns snapshot of the FLASH run as shown in Figure~\ref{fig:laserabs}. We match the profiles for electron density, flow speed, electron and ion temperatures, spatial coordinates in units of $d_{i0}=c/\omega_{\rm pi0}$ (proton skin depth; the subscript `$0$' denotes proton skin depth calculated for $n_e=n_{\rm cr}$) and temporal scales in terms of the hydrodynamical ion response time $d_{i0}/C_S$. In line with common practices in PIC simulations, we use a reduced speed of light (electron rest energy) and ion-electron mass ratio (increased electron mass) to accommodate the experimental timescale, typically on the order of the laser pulse duration (see Ref.\cite{Fox2018} and references therein for detailed examples and applications). In this work, we use a mass ratio of $m_p/m_e=100$ and $m_ec^2_\ast = 60 \,\rm keV$. 

The simulation box is one-dimensional, spanning $1000 d_{e0} = 100 d_{i0}$ and is resolved with 5000 grid cells. We use $10^5$ equally-weighted particles per cell per species at critical density. This particle number allows us to resolve densities greater than $10^{-5} n_{\rm cr}$. We do not introduce any chamber gas (as is required by the hydrodynamic approach) since we are unable to fully capture the physics at such low densities. In addition, the solid target is limited to a maximum density of $n_{\rm max,PIC} = 10 n_{\rm cr} \ll n_{\rm max,FLASH}$ to reduce the total particle count and ease the computational demand. The consequences of these numerical simplifications are addressed in Appendix~\ref{sec:convergence}. A CFL number of 0.75 is used to calculate the timestep and reflecting boundary conditions are used. The aluminum target is fully ionized and its thickness is $4.5 d_{i0}$. The initial location of the solid-vacuum interface is set to be at $z=0$ in both PSC and FLASH. The subscript `$0$' is omitted in subsequent references for brevity. Due to the reduced simulation parameters (reduced $m_p/m_e$, $m_ec^2$, and $n_{\rm max, PIC}$), a convergence study is necessary to validate the primary results discussed in this paper. Appendix~\ref{sec:convergence} details the findings of this convergence study. To briefly summarize the convergence studies, we find that the numerical parameters of reduced mass ratio, reduced speed of light, and reduced target density used in this work converge well with a range of simulations that have stricter parameters.

In PSC, collisions are implemented with the binary collision Monte-Carlo module\cite{Germaschewski2016}. Caution should be used when modeling collisions in PIC systems with reduced speed of light and mass ratio parameters, since they will modify the relative role of $e-e$, $e-i$, and $i-i$ collisions if treated uniformly. A special approach to match $e-i$ and $i-i$ collisional rates is implemented in PSC\cite{Totorica2024b}, allowing for a better retention of electron and ion heat transport properties. We currently apply a single value of the Coulomb logarithm to the whole simulation box, which is a simplification that may be improved in future work, but provides a first step and generates results that converge with FLASH simulations. The two most important collisional effects, namely, electron heat transport and electron-ion equilibration, were tested using PSC simulations with collisions and demonstrate good agreement with theoretical values across a wide range of plasma parameters. See Appendix~\ref{sec:eitest} and \ref{sec:thermaltest} for a detailed description of these tests.

Figure~\ref{fig:laserabs} illustrates the initial condition for the 1D PSC simulation, highlighting similarities and differences with the FLASH t\,=\,0.1 ns snapshot, and providing a benchmark for the PSC laser absorption module. We see the aforementioned differences in electron density profile and observe a very good agreement between FLASH and PSC laser power absorption profiles. We run a 1D PIC simulation starting from the snapshot depicted in Fig.~\ref{fig:laserabs} for 1 ns. In the next section, we will dive deeper into the evolution of plasma profiles in both PSC and FLASH, and various metrics related to extended hydro and kinetic effects.

\section{Results}\label{sec:results}

\subsection{Plasma profile evolution}

\begin{figure}[!ht]
    \centering
    \includegraphics[width=\linewidth]{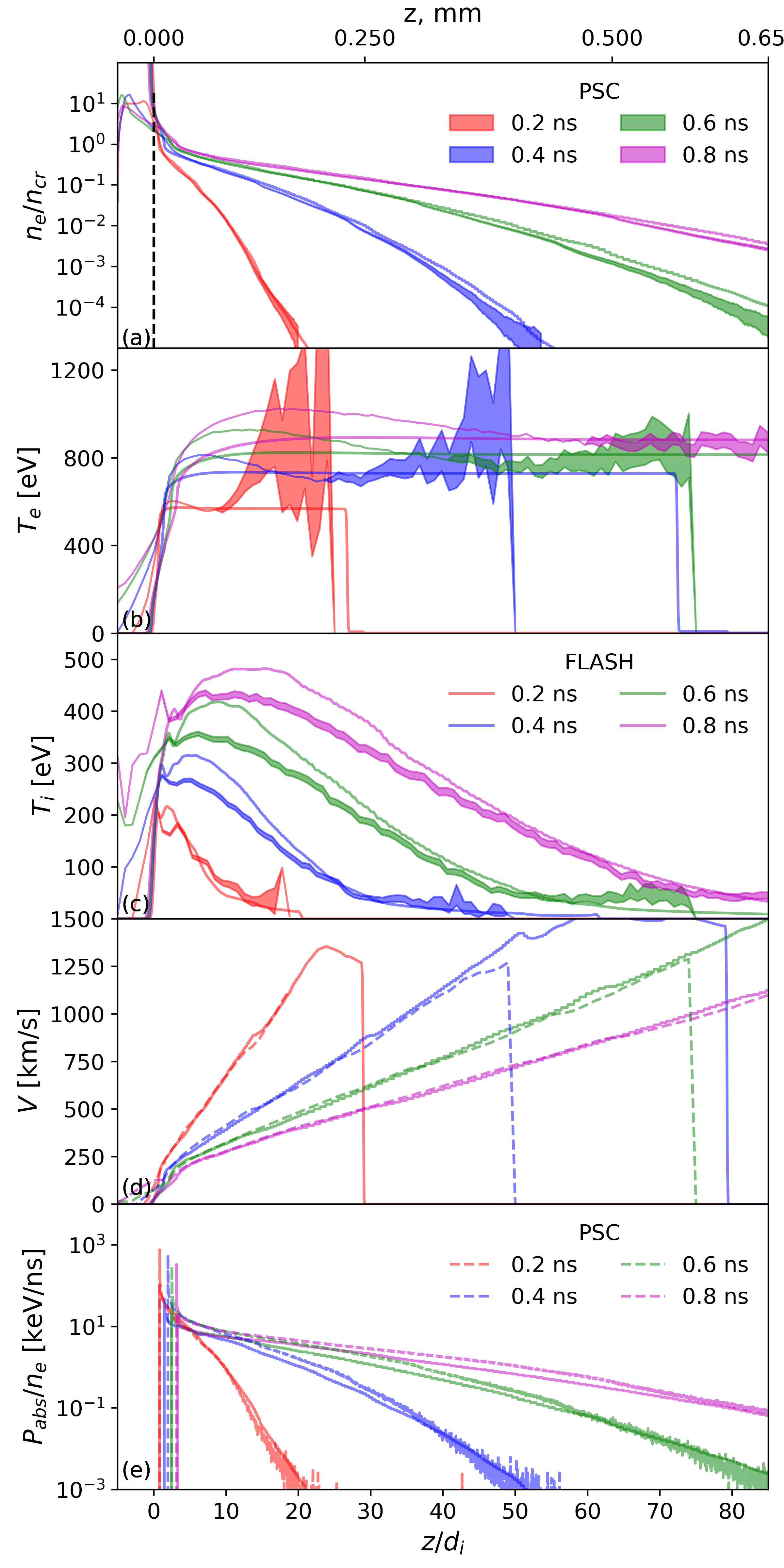}
    \caption{Comparison of plasma profile evolution between FLASH (solid lines) and PSC (shaded regions or dashed lines). (a) Electron density, (b) electron temperature, (c) ion temperature, (d) plasma expansion velocity, and (e) laser power absorption profiles. The vertical dashed line in (a) represents the target edge at $t=0$ which is initially located between $z=-4.5$ and 0 $d_{i}$. Good agreement between FLASH and PSC is observed.}
    \label{fig:flash_psc_evol}
\end{figure}

Let us start by comparing the first moments of the distribution function that are present in both FLASH and PSC (implicitly). Specifically, we consider the evolution of electron density, electron and ion temperatures, and flow speed. These four moments are present in FLASH and are evolved using the mass, momentum, and energy conservation equations (Eqs.~\ref{eqn:cont}-\ref{eqn:eele}). In FLASH, electron density is calculated from mass density as $n_{e,\rm FLASH} = \rho \bar{z}/m_{\rm Al}$, where $m_{\rm Al} = 26.98 m_p$ is the aluminum atomic mass and $\bar{z}$ is the average ionization state, determined from the tabulated EOS for the local conditions of $T_e,\, T_i$, and $\rho$. In this run, we consider a fully ionized target so that $\bar{z}=13$. 

In PSC, statistical errors introduced by the finite number of particle are estimated using the Poisson distribution for density and a bootstrap approach for temperatures and flow speed. 95\% confidence intervals are depicted with shaded regions. Moments of the electron and ion distribution function are calculated from the full particle data \cite{Fox2011, Fox2018}.

\begin{figure*}
    \centering
    \includegraphics[width=\linewidth]{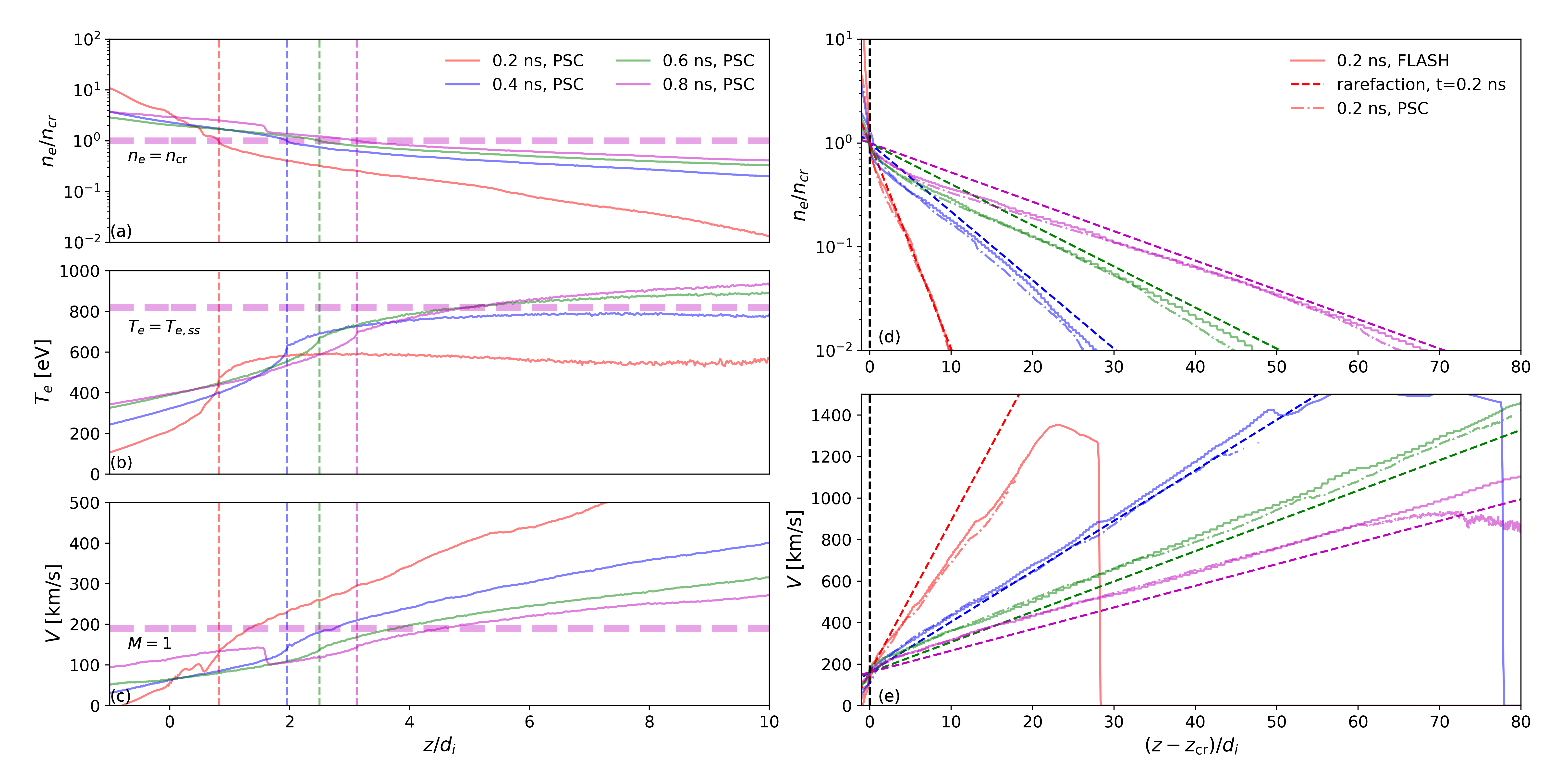}
    \caption{Analysis of critical surface dynamics, plasma thermalization, and expansion dynamics. The left panels show zoomed-in PSC lineouts (solid line) of (a) electron density, (b) electron temperature, and (c) flow speed at t=0.2, 0.4, 0.6, 0.8 ns. The location of the critical surface is shown in dashed vertical lines of the corresponding color. The steady-state (SS) model predictions for the critical density, electron temperature, and expansion speed are given by horizontal dashed magenta lines. The right panels show (d) electron density and (e) flow speed evolution at a larger spatial scale in PSC, FLASH, and the SS model. Good agreement is observed among the simulations and SS model.}
    \label{fig:mach_temp_rarefaction}
\end{figure*}

Figure~\ref{fig:flash_psc_evol} compares the evolution of expanding Al plasma profiles in PSC (shaded regions or dashed lines) and FLASH (solid lines) simulations. Note that the timestamps correspond to the FLASH simulation time and PSC is initialized at $t=0.1$ ns. Subfigure (a) shows the electron density profiles and demonstrates agreement to within 20\% in the underdense region where $n_e<n_{\rm cr}$. As expected, the density differs inside the solid target due to different initial density conditions between FLASH and PSC. The electron temperatures also agree within 20\% between the codes, although there are several distinguishing features. Firstly, the PSC electron temperature profile in Fig.~\ref{fig:flash_psc_evol}b is doubly-peaked. One peak is located near the critical surface and another is at the edge of the expanding plasma, near the limit of plasma density resolution. This contrasts with the isothermal plasma expansion observed in FLASH. Secondly, we observe a more gradual cooldown toward the solid target in PSC compared to FLASH (see Fig.~\ref{fig:flash_psc_evol}b, $z<0$), which implies the difference in thermal transport properties in PSC and FLASH, which we will discuss below. Ion temperatures also agree between FLASH and PSC (also within 20\%), with a slightly more spread-out ion temperature profile in PSC, possibly due to the absence of the ion heat flux in this FLASH run, see Eq.~\ref{eqn:eion}. The plasma flow speed profiles show agreement to within $<10\%$ and increase linearly in space, which is a common feature in isothermal plasma sheath expansion theories and simulations\cite{Manheimer1982,Mora2003,Hemminga2021}. Due to similar electron temperature and density profiles between the codes, the laser power absorption profiles also closely resemble each other, further validating the laser power absorption model in PSC\cite{Hyder2024}. Importantly, auxiliary simulations with either collisions or laser heating turned off demonstrated drastically different plasma evolution, highlighting the importance of both laser heating and collisional dynamics in the plasma expansion.

\subsection{Comparison against the steady-state solution}

Given the relative simplicity of the system, it is instructive to compare the simulations against an analytical solution of steady-state ablation. In the work by Manheimer et al.\cite{Manheimer1982}, a steady-state solution is found based on assumptions of mass, momentum, and energy conservation across the solid-plasma interface and critical surface, along with a rarefaction wave ansatz for expanding underdense plasma and a delta function-like laser absorption at the critical surface. They demonstrated that the ablation process solely depends on laser intensity, wavelength, and target material.

The main conclusions from their theoretical analysis may be summarized as follows.
(1) Expanding underdense plasma is isothermal, with electron temperature given by:
    \begin{equation}
        T_{e,{\rm SS}} = 5.94\, \mu^{1/3}Z^{-1/3} \left(\frac{\lambda}{1\,\rm \mu m}\right)^{4/3} \left(\frac{I}{I_{10}} \right)^{2/3}\; \textrm{eV},
    \label{eqn:temp_manheimer}
    \end{equation}
\noindent Here, $\mu$ is the mass ratio of ion to proton mass, $\lambda$ is the laser wavelength measured in microns, $Z$ is the ionization state,  $I$ is the laser intensity in $\rm W/cm^2$, and $I_{10} = 10^{10}\rm \, W/cm^2$. (2) Subcritical plasma expands according to the rarefaction wave ansatz:

    \begin{eqnarray}
        n_{e,{\rm SS}} = n_{\rm cr} \exp{(-z/C_St)}, \label{eqn:rarefaction1} \\
        V_{z,{\rm SS}} = C_S+z/t, \label{eqn:rarefaction2}
    \end{eqnarray}

\noindent where $C_S= \sqrt{Z T_{e,{\rm SS}}/m_i}$ is the sound speed corresponding to $T_{e, SS}$. (3)The critical surface location $z_{\rm cr}$, defined as the solution of the equation $n_e(z_{\rm cr}) = n_{\rm cr}$, moves in the laboratory frame (i.e., frame in which the solid target is initially at rest) with Mach one speed ($V_z(z_{\rm cr}) = C_S$). (4) The temperature inside the overcritical plasma layer scales with the spatial coordinate as:

    \begin{equation}
        T_{e}(z) = T_{e,\rm SS} \left(1+\frac{25}{4}\frac{n_{\rm cr}k_B^{3/2} T_{e,\rm SS}^{1/2}}{m_e^{1/2}K_e}(z-z_{\rm cr}) \right)^{2/5},
    \label{eqn:Teabl}
    \end{equation}
    
\noindent where $K_e$ is calculated via Eq.~\ref{eqn:Kele} at $T_e=T_{e,\rm SS}$. When z is small enough such that $T_e=0$, it is assumed that the solid-ablation layer interface is reached.

A comparison of PSC and FLASH simulations against the steady-state ablation model (hereafter referred to as the SS model) is helpful to benchmark simulations and understand whether the model assumptions are appropriate.

\begin{figure}
    \centering
    \includegraphics[width=\linewidth]{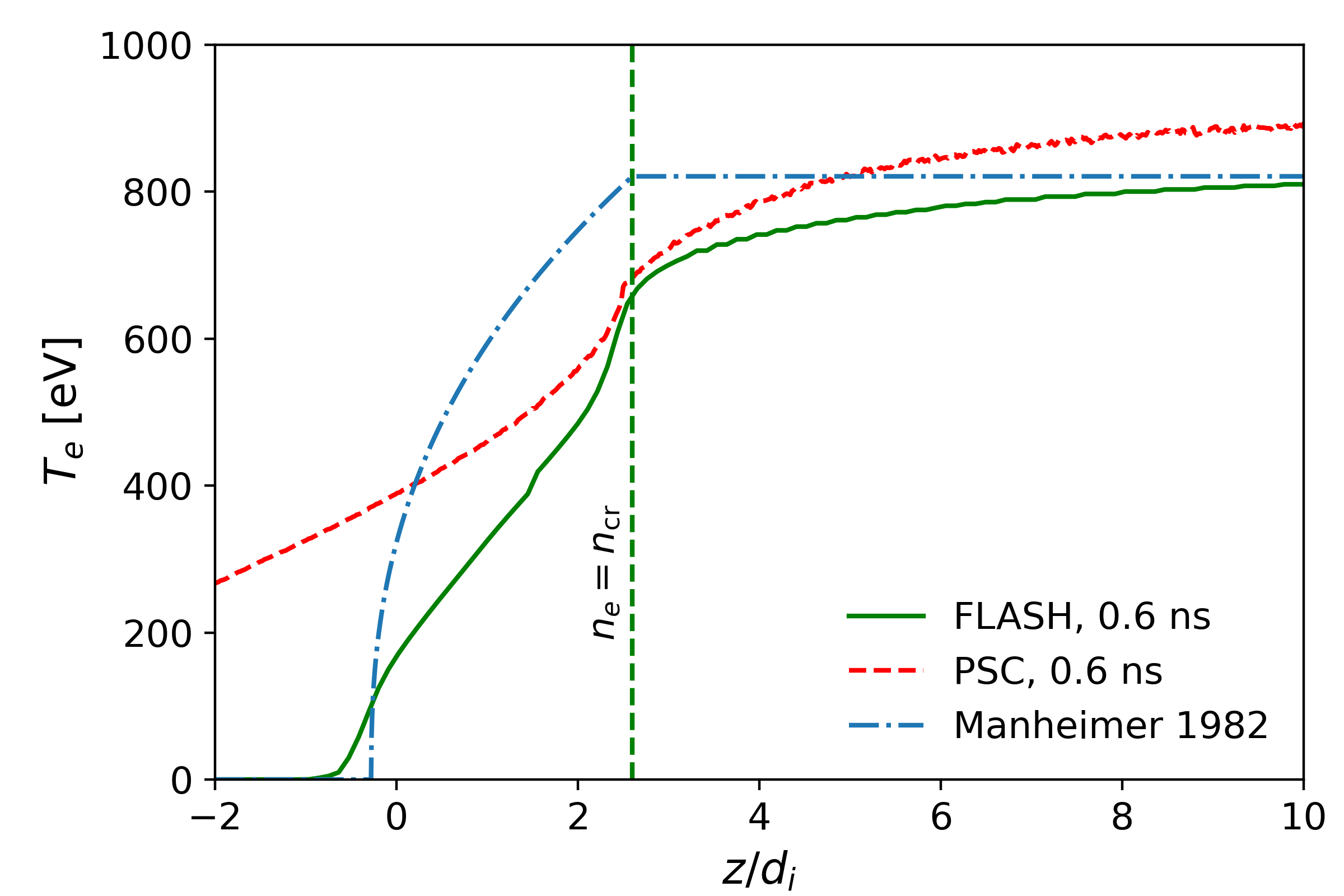}
    \caption{Electron temperature profiles close to the solid target surface at t=0.6 ns. FLASH (green solid), PSC (red dashed), and theoretical prediction given by Eqn.~\ref{eqn:Teabl} (blue dot-dashed) are displayed. The critical surface is denoted by the vertical dashed line.}
    \label{fig:heat_abl_front}
\end{figure}

\par
Figure~\ref{fig:mach_temp_rarefaction} presents an analysis of critical surface dynamics, electron temperature saturation, and the rarefaction wave solution. Figures on the left column illustrate profiles of (a) electron density, (b) electron temperature, and (c) flow speed close to the solid target (note that the z-axis is zoomed to 10 $d_i$). The solid color lines denote 1D PIC simulations at different times and the horizontal dashed magenta line shows the $n_e=n_{\rm cr}$ level in (a), the steady-state temperature given by Eq.~\ref{eqn:temp_manheimer} in (b), and Mach one at the same temperature in (c). Vertical dashed lines denote the location of the critical surface at different times. At t\,=\,0.2 and 0.4 ns, the plasma remains colder than the steady-state electron temperature. However at later times, the plasma is heated above this temperature, possibly due to the extended inverse Bremsstrahlung absorption region in the simulations. The Mach number evolution also suggests that the steady-state ablation regime is not reached during the simulation, with the critical surface moving at $V_z \leq 0.8 C_S$. Figures ~\ref{fig:mach_temp_rarefaction}d and e compare the rarefaction solution (Eqs.~\ref{eqn:rarefaction1}-\ref{eqn:rarefaction2}) against FLASH and PSC at different times. The rarefraction profiles nominally agree with the PSC and FLASH behavior in terms of density blowoff (d) and flow speed (e). Note that rarefaction profiles are calculated for the sound speed $C_S$ observed in the simulations at the critical surface, which is less than the SS temperature (0.8$T_{e,SS}$).

Interestingly, heat transport inside the ablation front (between the solid-plasma interface and $n_e=n_{\rm cr}$) behaves differently in all three approaches. Figure~\ref{fig:heat_abl_front} compares electron temperature profiles in FLASH, PSC, and SS models at t\,=\,0.6 ns. While the SS model and FLASH predict a similar location for the solid-ablation layer interface, their temperature profiles inside the ablation layer differ. This could be attributed to the heat flux limitation in FLASH, which reduces the heat flux around the critical surface and steepens the temperature profile at $z=z_{\rm cr}$. On the other hand, PSC has a much flatter temperature profile due to the capped $n_{\rm max}$ in PIC simulations. However, as $n_{\rm max}$ is increased in convergence studies, PSC temperature profiles tend to align closer with FLASH and SS predictions, as demonstrated in Appendix~\ref{sec:convergence}.

\subsection{Heat transport in FLASH, PSC, and SS model}

\begin{figure*}
    \centering
    \includegraphics[width=\linewidth]{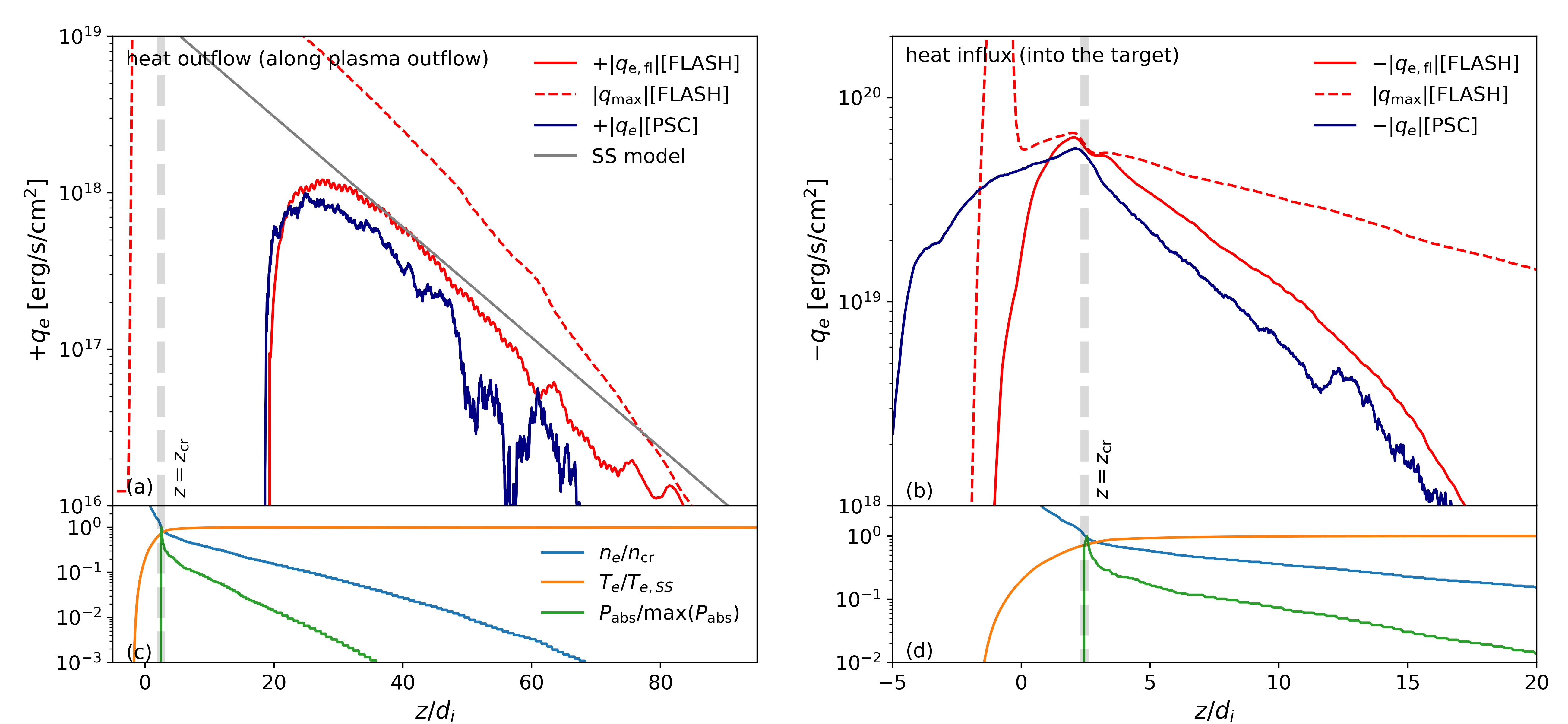}
    \caption{Heat fluxes calculated for FLASH, PSC, and the SS model at t=0.6 ns. Both outflowing (along the plasma expansion, (a)) and inflowing (toward the solid target, (b)) heat fluxes are depicted on separate subplots. Electron density (blue), temperature (orange), power deposition profile (green) are shown for reference in subplots (c) and (d). Good agreement is seen among all three models in the $+z$ direction, and notable differences are observed between FLASH and PSC in the $-z$ direction. In FLASH, $|K_{\rm ele}\nabla T_e|$ approaches $q_{\rm max}$ inside the target, which results in a flux-limited $q_{\rm e,fl}$.}
    \label{fig:heat_flux}
\end{figure*}

To verify that we correctly capture the heat fluxes around the critical surface and in the underdense plasma, we compare the electron heat fluxes in FLASH, PSC, and the SS model. In FLASH, the classical Spitzer heat flux is given by Eqs.~\ref{eqn:heatflux} and \ref{eqn:Kele} and the thermal conduction is flux-limited (see Section~\ref{sec:setup} for details). In PSC, the heat fluxes are calculated directly from the particles: $q_{ez,i} = m_ec^2 \sum_{j: z_j \in (z_i,z_i+dz)}w_jv_{jz}'(\gamma_e'-1)$. Here, $w_j$ is the particle weight (density each particle contributes; PSC uses an equal weight for all particles of the same kind), $V_{\rm flow}$ is the electron flow speed in the cell $i$, $\vec{v_{j}}'$ is the electron velocity in the frame moving with $V_{\rm flow,i}$, $\gamma_e' = 1/\sqrt{1-|\vec{v_{j}}'|^2/c^2}$. Reference frame transformations are done using a Lorentz transformation of velocities. The equation for $q_{ez,i}$ yields the expected expression for the heat flux in the non-relativistic regime.

Finally, the heat flux can be calculated in the SS model by considering the equation for electron energy evolution (Eq.~\ref{eqn:eele}) without the laser heating, e-i thermalization, or time-dependent terms. This simplifies into $P_e \partial V_z/\partial z= - \partial q_{ez}/\partial z$ and leads to an outgoing heat flux of $q_{ez} = n_eT_{e,\rm SS} C_S\rvert_{T_e=T_{e,\rm SS}}$. The spatio-temporal dependence in $n_e$ is given by Eq.~\ref{eqn:rarefaction1}, in agreement with Eq.~10 from Ref.\cite{Manheimer1982}. Ignoring the ion contribution in the heat flux calculation is justified, since it is smaller by a factor of $1/Z$, and $T_e>T_i$ for PSC and FLASH simulations.

\begin{figure*}
    \centering
    \includegraphics[width=\linewidth]{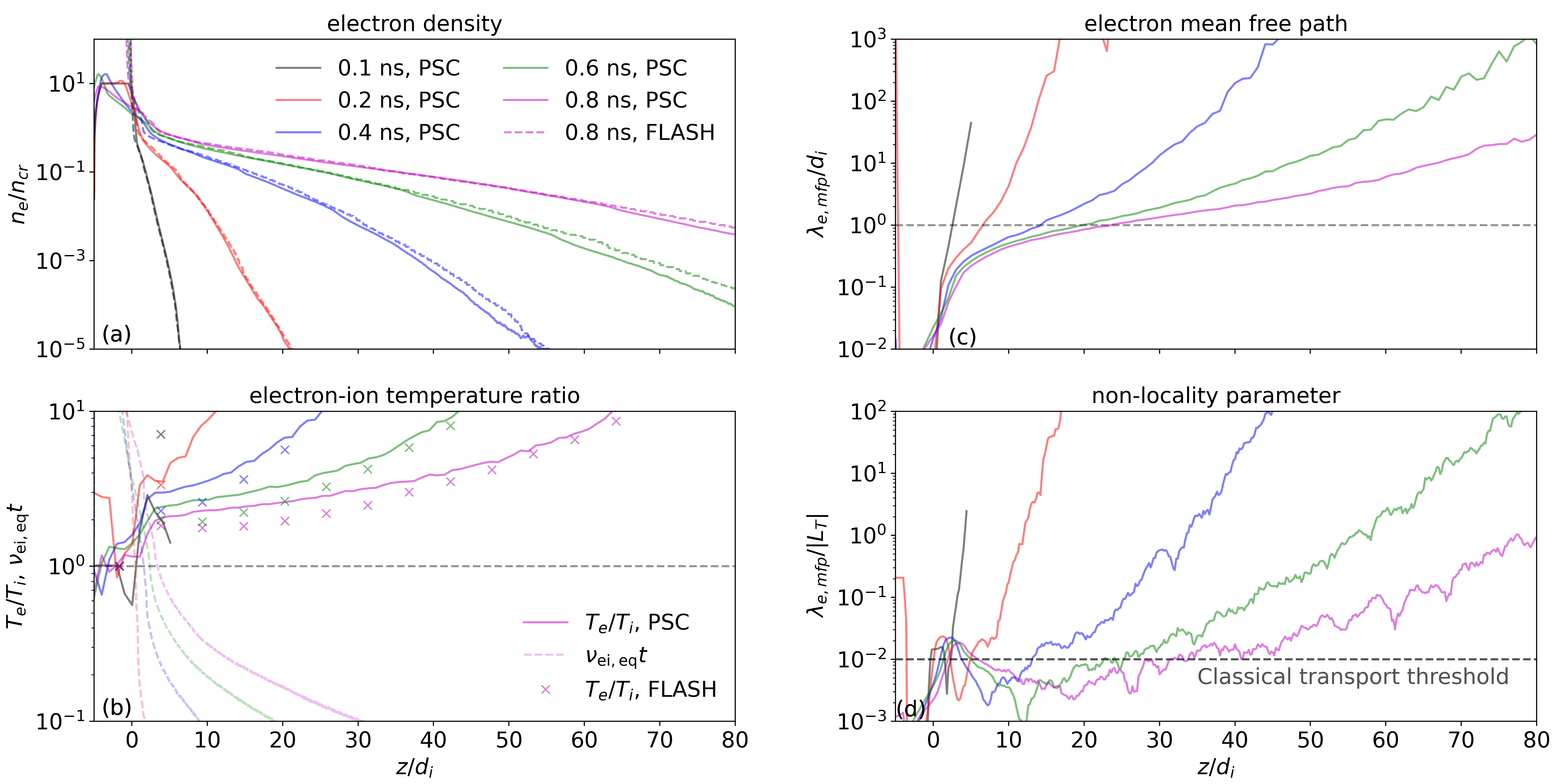}
    \caption{Kinetic effects in PSC. (a) Electron density profiles. (b) Comparison of electron-ion temperature ratio between PSC and FLASH, $T_{e}/T_{i}$ and equilibration rate, $\nu_{\rm ei,eq}t$. (c) Electron mean free path normalized to $d_{i}$ (i.e., proton skin depth at $n_e=n_{\rm cr}$). (d) Non-locality parameter (electron mean free path normalized to local temperature gradient, $L_T$).}
    \label{fig:kinetic_effects}
\end{figure*}

Figure~\ref{fig:heat_flux} compares these models at t=0.6 ns, both for heat fluxes with positive sign (heat flowing along the plasma expansion in the $+z$ direction, Fig.~\ref{fig:heat_flux}a) and negative sign (heat flowing into the solid target in the $-z$ direction, Fig.~\ref{fig:heat_flux}b). Due to the flux-limiting in FLASH, both the limiting flux, $q_{\rm max}$, and the resulting FLASH heat flux, $q_{\rm e,fl}$, are plotted for clarity. Additionally, we depict the electron density, temperature, and deposition profiles on the bottom subplots for reference. The positive heat fluxes agree within 30\% between all three models. Similarly, the negative heat fluxes in FLASH and PSC agree for $z>z_{\rm cr}$ to within a factor of 2 of each other. However, they also exhibit qualitative differences, especially in the solid target region. Here, the heat flux in PSC extends deeper into the solid slab, which results in a more diffusive electron temperature profile in PSC. This diffusive behavior is expected, since the solid density in PSC has been artificially reduced, thus reducing plasma collisionality and amplifying hot electron diffusion in both the $+z$ and $-z$ directions. The discrepancy in negative heat fluxes can also be attributed to the calculation of $\ln \Lambda_{ei}$ in FLASH and PSC -- the former uses a local value calculated from Eqs.~\ref{eqn:lnl1}-\ref{eqn:lnl3}, whereas PSC adopts a global value based on a characteristic plasma density, temperature, and ionization without accounting for local variations in $n_e$ and $T_e$. Nevertheless, PSC captures the underdense plasma expansion well and agrees with FLASH simulations to within 20\%. Finally, the FLASH heat flux is close to the flux-limited value for $-z$ heat fluxes around the critical surface. The resulting FLASH heat flux is within a factor of two from the PSC estimates inside the ablation front, as shown in the blue line in Fig.~\ref{fig:heat_flux}b.

One notable difference between the SS model and both the PSC and FLASH simulations is the location where the thermal conduction heat flux reverses sign, $q_{ez}(z_{\rm hr}) = 0$. In the SS model, this point is assumed to be at the critical surface, however the reversal point happens much further away for both FLASH and PSC. At t=0.6 ns, the critical surface is $z_{\rm crit} \approx 3 d_{i}$, but the heat flux sign reversal happens between $z/d_{i}=15$ and $20$. This discrepancy may be explained by the spatially extended laser energy deposition, which remains significant (on a level of $P_{\rm abs}>10^{-2}P_{\rm abs,\, max}$) further away from $z_{\rm crit}$. This effectively shifts the location where the laser deposition vanishes, which is where the sign reverses in Manheimer et al\cite{Manheimer1982}.

\subsection{Deviations from fluid behavior in PSC}

Thus far, we have demonstrated convergence between FLASH and PSC in terms of underdense plasma behavior and highlighted differences in thermalization within the overcritical ablation front. Yet one of the main points of using PIC simulations to model plasma ablation is to capture extended hydrodynamic and kinetic effects that hydrodynamic simulations may miss. With that in mind, this section details the kinetic physics that PSC is able to detect that goes beyond the hydrodynamic model.

Some hydrodynamic codes, including RALEF \cite{RALEF}, employ a single-fluid single-temperature model, which is applicable to the regime of smaller laser intensities (e.g., $10^{9}-10^{11} \,\rm W/cm^2$ used in extreme ultraviolet lithography sources\cite{Hemminga2021}), but may not be suitable for HED plasmas with laser intensities of $10^{13} \rm \, W/cm^2$ and above. These plasmas have large $T_e/T_i$ ratios and do not equilibrate on the timescale of the pulse duration. Other codes, including FLASH, implement a single-fluid two-temperature model, allowing for $T_e \neq T_i$ and incorporating the e-i collisional equilibration term in the energy equations (Eqs.~\ref{eqn:eion},\ref{eqn:eele}). Comparing the $T_e/T_i$ profiles is therefore a good benchmark for FLASH and PSC and details regarding the e-i thermalization testing in PSC may be found in Appendix~\ref{sec:eitest}. 

For hot ($\sim 1 \rm \, keV$) and tenuous ($n_e<n_{\rm cr}$) plasmas expanding supersonically ($V\sim 10^3 \rm \, km/s$), the long mean-free-path effects begin to influence the plasma dynamics. We estimate the mean free path of electrons, $\lambda_{\rm e,mfp}$, and  compare it to both the typical spatial scale of an HED experiment, the ion skin depth, as well as to the lengthscale of the electron temperature gradient. Classical heat transport theory assumes that $\lambda_{\rm e,mfp}/L_T \ll 1$ (see, e.g., Ref.~\cite{Mora1982}), with $L_T \equiv |T_e/ (dT_e/dz)|$ which may be violated in certain HED plasmas\cite{Bell1981}. Nonlocal transport models have been developed to address the issue\cite{Holec2018}, as the classical heat transport model is unable to reproduce the correct heat flux values, even with flux limiting\cite{SNB}. Despite these advances, certain kinetic physics may still be missing\cite{Sherlock2017}, which motivates additional kinetic modeling to further improve heat transport in HED.

Figure~\ref{fig:kinetic_effects} shows several metrics where the PSC simulation may depart from a fluid behavior. First, we present electron density profiles, Fig.~\ref{fig:kinetic_effects}a, which are the same as in Fig.~\ref{fig:flash_psc_evol}, for reference. Figs.~\ref{fig:kinetic_effects}b,c,d are devoted to collisionality-dependent effects. Figure~\ref{fig:kinetic_effects}b compares electron-ion temperature ratio $T_e/T_i$ profiles between FLASH and PSC and shows good agreement between them. Theoretical equilibration rates normalized to the simulation time, $\nu_{\rm ei,eq} t$, are plotted as dashed lines and shown to be slow for underdense plasma, thus highlighting the importance of treating temperatures separately for plasma ablation in the HED regime. The electron mean free path profile is plotted in Fig.~\ref{fig:kinetic_effects}c, normalized to the ion skin depth, and in Fig.~\ref{fig:kinetic_effects}d, normalized to the local temperature gradient length scale. Sufficiently far from the target, the electron mean free path becomes comparable to the local temperature gradient scale (i.e., within a factor of 10). Ref.\cite{Matte1982} suggests that deviations from classical transport happen at $\lambda_{\rm mfp}/L_T \gtrsim 10^{-2}$ at near-critical density close to the peak of laser energy deposition. Preserving the correct electron thermal transport physics there is important for the analytical prediction of the ablating plasma properties, and, thus, a model beyond Spitzer may be necessary to incorporate such effects \cite{Holec2018}.

In addition, any future interaction with a background or counterstreaming plasma should be treated kinetically and result in plasma interpenetration and anisotropy dissipation via collisionless mechanisms. Interactions such as these are routinely modeled with kinetic simulations in support of experimental campaigns involving one or multiple supersonically expanding plasma plumes\cite{Fox2018,Fiuza2020,Lezhnin2021,Grassi2021}. 

\begin{figure}
    \centering
    \includegraphics[width=\linewidth]{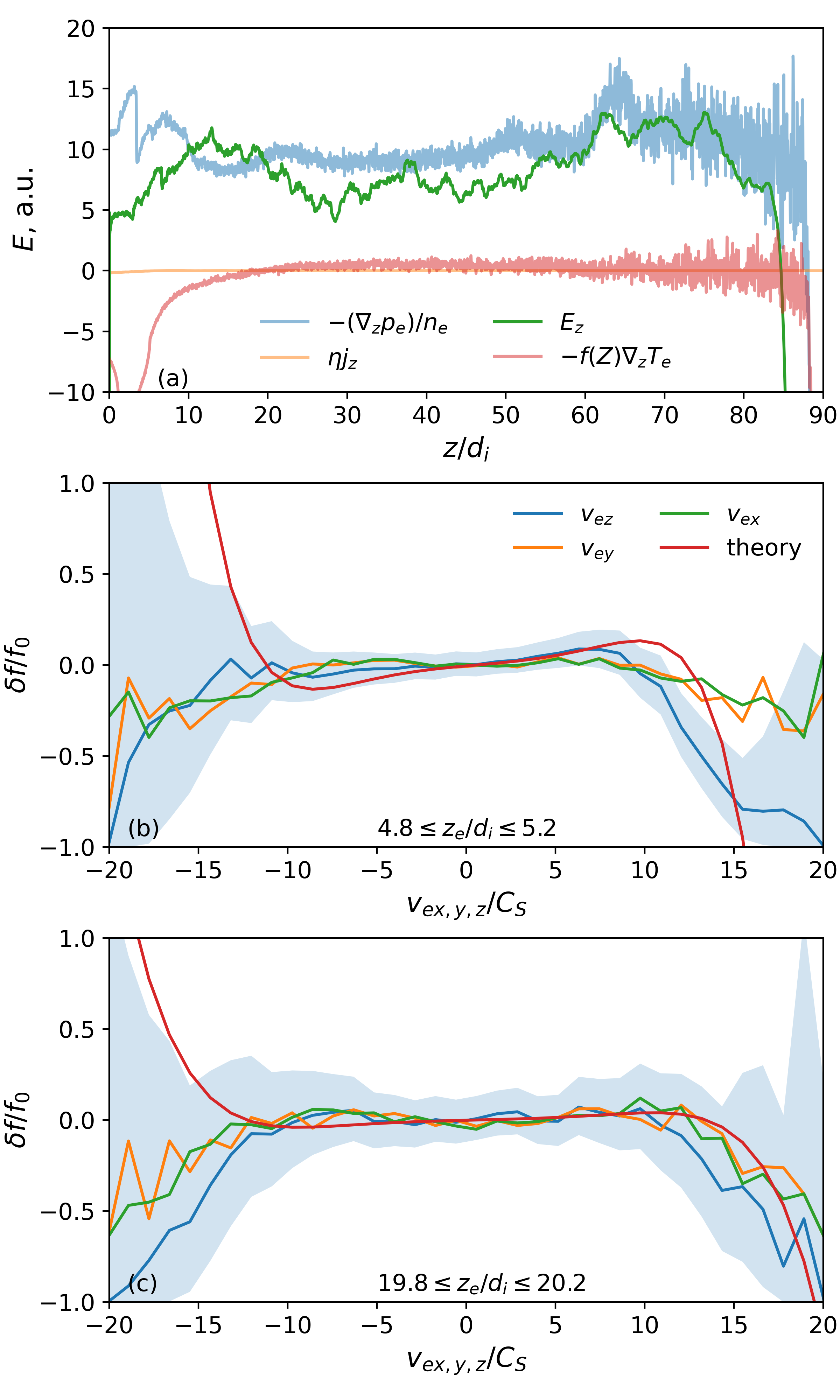}
    \caption{(a) Generalized Ohm's law and its main contributors in arbitrary units. (b) Relative perturbation of electron velocity distribution functions along x,y,z at $4.8 \leq z/d_i \leq 5.2$ and (c) $19.8 \leq z/d_i \leq 20.2$ with respect to local Maxwellian. All quantities are measured from a PSC snapshot at t=0.8 ns. Shaded region represent histogram estimator 95\% confidence interval of $f(v_z)$, and red line represents distribution function perturbation due to the presense of thermal heat flux from classical Spitzer heat conduction theory.}
    \label{fig:pdf_ohm}
\end{figure}

To assess the plasma model under consideration, we calculate the terms of the generalized Ohm's law from the PIC simulations and their relative contribution. Figure~\ref{fig:pdf_ohm}a compares electric field components and the compensating terms along the expansion direction at t=0.8 ns. The dominant terms are the electric field and the electron pressure term $-(\nabla p_e)/n_e$, while the resistive term $\eta J$ and the electron inertia term are negligible. The thermal force $-f(Z)\nabla_z T_e$ (see Ref.~\onlinecite{Braginskii}) is also small in the underdense plasma but does have some effect in the solid target ($f(Z)$ is the function of mean ionization which tends to $1.5$ at $Z \gg 1$). The resulting Ohm's law, $E = - (\nabla p_e)/n_e$, is generally consistent with the hydrodynamic model. Still, the thermal force is important close to the critical surface, and should therefore be included in extended hydrodynamic closure for ablation simulations.

Finally, we may compare the velocity distribution function in PSC to the Maxwellian that is assumed in hydrodynamic simulations. Figures~\ref{fig:pdf_ohm}b,c compare relative perturbations of electron velocity distributions with respect to the local Maxwellian, $f_0 = (m_e/2\pi T_e)^{1/2}\exp{(-m_ev_{x,y,z}^2/2T_e)}$, along all three axes at two locations -- for electrons at (b) $4.8 \leq z/d_i \leq 5.2$ (close to the peak of temperature gradient and laser absorption) and (c) $19.8 \leq z/d_i \leq 20.2$ (where temperature gradient and laser absorption are less pronounced). We depict 95\% confidence interval of the histogram estimator of $f(v_z)$ with shaded region, as well as the Maxwellian distribution correction for the electron heat flux under the classical Spitzer heat conduction model, see Ref.~\onlinecite{Shkarofsky1966}:
\begin{equation}
    \frac{\delta f}{f_0} = \frac{1}{\sqrt{2\pi}} \frac{m_e^2v^4}{T_e^2} \left(8-\frac{m_ev^2}{T_e}\right)\frac{\lambda_{\rm mfp}}{L_T}.
\end{equation}
\noindent There is a clear deviation in the positive, z-directed velocity distribution (blue line) from a single-temperature Maxwellian at $z/d_i \approx 5$ (there, $n_e/n_{\rm cr}\approx 0.7$, $T_e \approx\, \rm 800\,eV$), while both transverse distributions are close to Maxwellian. The classical heat flux correction of the distribution function does not explain the observed $\delta f$ well. Interestingly, when we analyze the contribution of different speeds from electron distribution to the heat flux, we find that the dominant population of heat flux carrying electrons has $v_{ez}\approx 13 C_S$, which is close to where the $v_{ez}$ spectrum deviated from a Maxwellian. Further from the target, at $z/d_i \approx 20$, such strong deviation from a Maxwellian disappears. {The apparent non-Maxwellian feature close to the critical density further indicates the importance of non-local effects for thermal transport.} 

\section{Discussion}\label{sec:discussion}

In this paper, we have compared two approaches for modeling laser-driven solid target ablation in the high energy density physics regime. In the first approach, we used the radiation hydrodynamics code FLASH, while the second approach was a particle-in-cell simulation with laser ray tracing implemented in the code PSC. In general, the plasma profiles from both codes agree with each other and the steady-state ablation model, with minor differences due to numerical limitations of the PIC approach. However, the main discrepancy between the three is the temperature profile in the ablation front (i.e., in the overcritical plasma between the critical surface and solid-plasma interface). Still, PSC and FLASH heat fluxes in the underdense plasmas and in the ablation front are consistent with each other within the systematic uncertainty due to the Coulomb logarithm. There are several key takeaways from our PIC simulations in regards to possible long mean-free-path effects: (I) two-temperature physics is needed to accurately model the HED regime, since e-i equilibration for underdense plasma (including a near-critical one) is slower than the laser pulse duration; (II) electron mean free path calculations suggest collisionless behavior of the expanding underdense plasma and suggests the importance of non-local transport effects in the near-critical region; and (III) estimates of the generalized Ohm's law confirm the applicability of the simple fluid closure, although the electron distribution function analysis clearly indicates the persistance of anisotropy in the near-critical region, contributing to the heat flux suppression. These findings are important steps in the benchmarking of PSC as a code for HED physics, and highlight possible applications where PSC may provide insight into purely kinetic effects or construct closures to be implemented in hydrodynamic codes.

Below, we discuss the limitations of the PSC model, explore possible workarounds, and outline future steps for expanding the code. As shown in Figure~\ref{fig:heat_abl_front}, there are notable differences in heat transport near the solid target between PSC and FLASH. The discrepancies in PSC may be attributed to several factors: (a) reduced solid density to maintain a computationally reasonable number of particles within the solid target, (b) strict requirements on the collisional module regarding the frequency of collisional routine calls, (c) and self-heating stemming from numerical noise in the initially cold solid target. Issues (a) and (c) are typical limitations in PIC simulations, which may be mitigated by increasing resolution at the cost of higher computational expense. Similarly, increasing the frequency of collision routine calls from once every 10 PIC timesteps (base level of collision routine calls used in the paper) to once every PIC timestep would improve the accuracy of the collision module without dominating the computational load. Yet, simulating very dense ($n_e\sim n_{\rm cr}$) and cold ($T_e \sim 10^1 \rm \, eV$) plasmas on a long timescale ($>10^4$ timesteps) is computationally unfeasible for the collision module, as evidenced by e-i equilibration and thermal conductivity testing in Appendix~\ref{sec:eitest}. With the current capabilities of the PIC approach, accurately simulating thermal transport in solid regions over relevant timescales remains challenging. However, as demonstrated in Figure~\ref{fig:heat_flux}, PIC still adequately captures heat transport in near-critical and undercritical plasmas. Thus, our predictions about the plasma outflows are not affected by a more diffusive behavior of the PIC model in solid regions.

A few comments about the current implementation of the laser power deposition module should be made. First, the current work uses a simple inverse Bremsstrahlung absorption model to allow for direct comparisons against the FLASH code. However, recent experiments by Turnbull et al.\cite{Turnbull2023} have demontrated the need to include additional physics to correctly calculate the laser absorption rate. We can adapt the laser absorption model in PSC; for example, it was modified in Ref.\cite{Totorica2024} to match the RALEF-2D absorption model relevant for EUV source plasmas. Another important aspect of laser heating in PIC is the algorithm used to heat electrons according to the local laser power deposition rate. Currently, we implement heating by giving normally distributed random momentum kicks to the particles according to the local laser power deposition rate per particle. This approach inherently assumes a local Maxwellian particle distribution. However, the Langdon effect suggests that laser absorption via IB may lead to a super-Gaussian electron distribution function, which modifies absorption efficiency. Consequently, a theoretical closure for velocity-dependent laser heating is needed. Encouragingly, recent experimental and theoretical efforts, such as those by Milder et al.\cite{Milder2020,Milder2021}, offer potential models that could directly integrate into the laser heating routine in PSC.

One of the challenges in simulating target ablation is the ``initial plasma problem" \cite{Zhao2023} -- the sharply defined profile of the solid density target typically does not resolve undercritical densities on the numerical grid. This leads to situations where either $n_e \gg n_{\rm cr}$ or $n_e =0$ along a ray trajectory, resulting in a full reflection under the assumptions of simple laser ray tracing and power deposition models. In reality, however, the laser heats the electrons on the surface of the target which initiates the collisional ionization process and triggers plasma ablation. In FLASH, the ``slow start" approach is used to avoid this issue. The simulation is started with a very small timestep, which exaggerates the numerical diffusion and eventually creates an ablation layer where laser ray tracing and the IB model can be successfully applied. In PSC, we currently bypass this issue by initializing the simulation from a FLASH snapshot with an extended plasma profile. In principle, a similar result may be achieved by  depositing all the laser energy to the outermost layer of the target, effectively overriding the laser power deposition routine during the first few hundreds of timesteps. This method will be evaluated in future iterations of the PSC laser deposition model.

The current multi-dimensional version of the laser heating module in PSC assumes non-refracting and specularly reflecting rays propagating along one of the axes of the PIC numerical grid. While this approach limits the target geometries we can model, its validity is supported by comparisons with FLASH simulations. These comparisons demonstrate that the impact of refraction and reflection diminishes once laser absorption nears 100\%, which is common for laser-target interactions in the HED regime. A similar argument is used to justify the validity of the oblique incidence approach in our multi-dimensional simulations. To account for oblique angles of incidence, we modify the laser absorption coefficient\cite{Shearer1971} and maintain the assumption that the incident and reflected rays propagate along one axis of the simulation box -- a good assumption when nearly all of the incident ray's energy is absorbed, as detailed in Ref.\cite{Hyder2024}. This treatment is validated through auxiliary simulations that find comparable laser absorption at oblique incidences between PSC and FLASH simulations.

Future extensions to the PSC model will include dynamic ionization physics and radiation transport. Implementing dynamic ionization would enable calculations of mean ionization, which is a critical parameter for many HED plasma processes including laser absorption, thermal and radiation transport, electron conductivity, and electron-ion equilibration. Mean ionization is also important for applications centering around the radiation signatures of the laser-irradiated high Z solids, including radiation sources for lithography\cite{Versolato2019}. In these systems, ionization affects the radiation spectrum and may have complex dependences on the temperature and density of plasma along the line of sight. Currently, we treat mean ionization as a free parameter in our simulations and apply it uniformly to all ions to investigate its effect on the plasma dynamics.

Radiation transport is also crucial for simulations of EUV sources for lithography\cite{Versolato2019} and target interactions with x-ray lasers\cite{Sentoku2014}. Radiation transport was previously implemented in a particle code by Sentoku et al.\cite{Sentoku2014}, solving the radiation transport equation with an adaptive photon energy group selection and tabulated absorption and emission coefficients produced by the FLYCHK code \cite{FLYCHK}. Implementing these physics models into PSC will enable simulations of beyond-EUV sources that utilize higher laser intensities than the current state-of-the-art EUV source. These sources have higher electron temperatures ($\sim 10^2$ eV) and generate fast Gd ions of around $30\, \rm keV$ energy \cite{Niinuma2023}. 

We envision numerous applications of the PSC model in the field of HED physics. Since PIC codes are well-suited for treating plasma interpenetration, we can simulate experiments of laser-driven magnetic reconnection and collisionless shocks from first principles. Previous simulation attempts\cite{Fox2011,Lezhnin2018} used initial conditions based on assumptions about on plasma profiles without laser heating. Although this setup simplified the interpretation, it also made limiting assumptions about the distribution function in the expanding plasma plumes, which may modify the dynamics of the current sheet formation in the reconnection setup\cite{Fox2021}. Another aspect of the laser-driven reconnection experiment is the generation of magnetic fields via the Biermann-battery mechanism. Recent studies demonstrated the limited applicability of extended MHD models to predict the Biermann fields\cite{Ridgers2020}, suggesting the importance of capturing kinetic physics. Kinetic simulations with laser ray tracing can significantly contribute to solving this issue and serve as a benchmark for extended hydrodynamics closures for the Biermann problem\cite{Pilgram2022}. In the collisionless shock case, PIC simulations are initialized either from the idealistic two-plume collision setup\cite{Stockem2014}, by applying the volumetric heating operator to the solid target\cite{Fox2018,Schaeffer2020,Lezhnin2021}, or by starting a PIC run from 
a hydro simulation snapshot. In all three cases, we may miss details about the distribution function in the expanding plasma plume, thus suppressing the expected laser-plasma interactions or amplifying unphysical instabilities. The more rigorous method of laser heating via the ray-tracing module could account for the complexity of the plume formation and initial stages of the shock structure development.

Finally, one of the pinnacles of laser-driven radiation sources, the EUV lithography source, may also be simulated using PIC with ray tracing. Totorica et al.\cite{Totorica2024} demonstrate that PIC simulations can address fast ion debris formation\cite{Versolato2019,Hemminga2021}, a primary challenge in EUV lithography. Their 1D PIC simulations with laser ray tracing show that fast tin ions are accelerated up to a few keV in energy, primarily driven by electron pressure gradients. While these results generally agree with radiation hydrodynamic simulations by RALEF-2D, they demonstrate that the fast ion debris behave differently in kinetic simulations and hydro codes underestimate the fast populations by an order of magnitude or more. Future developments of the PSC model will improve the capabilities of EUV lithography source simulations, potentially addressing the issue of fast ion debris mitigation\cite{Israeli2023}.

To conclude, we show that PSC with laser ray tracing is a complementary approach to the radiation hydrodynamic simulations in the HED regime. This method is particularly well-suited for addressing anisotropies, plasma interpenetration, and benchmarking extended hydrodynamic models.

\section*{Data Availability Statement}
The data that support the findings of this study are available from the corresponding author upon reasonable request. 

\section*{Acknowledgements} 

This work was supported by the U.S. Department of Energy under contract number DE-AC02-09CH11466. This work was supported by the Laboratory Directed Research and Development (LDRD) Program of Princeton Plasma Physics Laboratory. This paper describes objective technical results and analysis. Any subjective views or opinions that might be expressed in the paper do not necessarily represent the views of the U.S. Department of Energy or the United States Government. Sandia National Laboratories is a multi-mission laboratory managed and operated by National Technology \& Engineering Solutions of Sandia, LLC, a wholly owned subsidiary of Honeywell International Inc., for the US DOE’s National Nuclear Security Administration (NNSA) under contract DE-NA0003525. The simulations presented in this article were performed on computational resources managed and supported by Princeton Research Computing at Princeton University. The Flash Center acknowledges support from the US DOE NNSA under awards DE-NA0004144 and DE-NA0004147, and subcontract No 60138 with Los Alamos National Laboratory. The software used in this work was developed in part by the DOE NNSA- and DOE Office of Science-supported Flash Center for Computational Science at the University of Chicago and the University of Rochester.

\appendix

\section{Convergence tests}\label{sec:convergence}

To verify the validity of the PIC simulation results discussed throughout the paper, we conducted a series of convergence tests. We varied the maximum initial density in the solid target regions in PIC simulations, $n_{\rm max, PIC}$, from $2 n_{\rm cr}$ to $20 n_{\rm cr}$, while keeping the same $N_{\rm PPC}=10^5$ at $n_e=n_{\rm cr}$ for all runs except for the one with $n_{\rm max,PIC}=20n_{\rm cr}$, where $N_{\rm PPC}=10^4$ was used. For the reference, nominal electron density for the fully ionized solid aluminum target is $n_{\rm e,solid}\approx 700 n_{\rm cr}$. We also varied the parameter $m_ec_\ast^2$ value from 20 to 200 keV, which reflects the reduced speed of light in the simulation. When $T_e \ll m_ec^2_\ast$, the electrons will remain non-relativistic. The proton to electron mass ratio was also scanned from the text value of 100 to an additional value of 400 here. For this run, we kept the resolution and the box size in $d_i$ units the same as in the primary simulation ($5$ grid cells per $d_{e}$ and $100 d_{i} = 2000 d_e$) and considered a reduced $m_ec_*^2$ value of 20 keV to relax the computational requirements of the simulation.

\begin{figure}
    \centering
    \includegraphics[width=\linewidth]{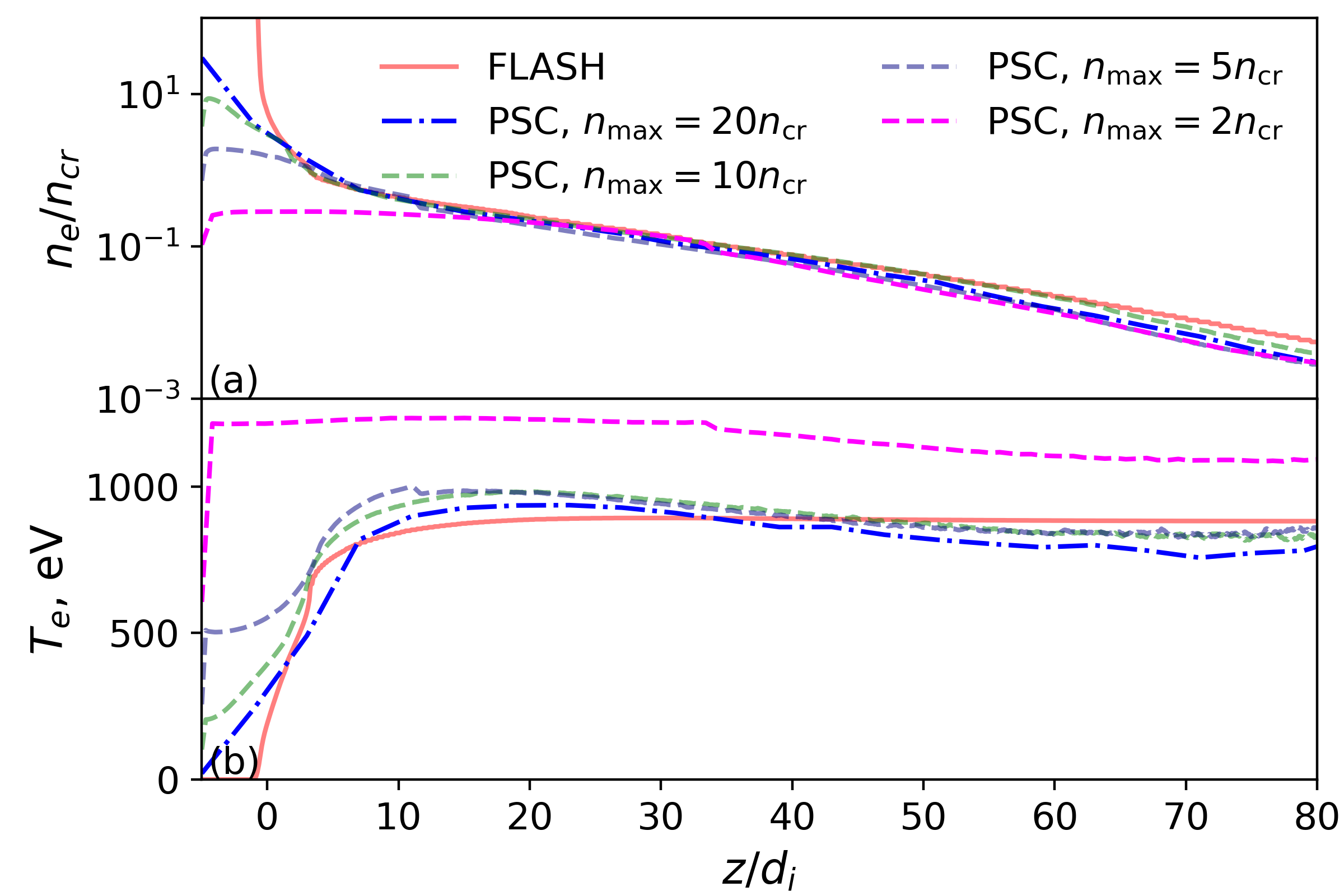}
    \caption{Convergence test for the solid target density used in PSC, $n_{\rm max}$. The densities considered are 2, 5, 10, 20 $n_{\rm cr}$, shown in dashed lines, and compared to a FLASH simulation in the solid red line. (a) Electron density and (b) electron temperature profiles. The underdense plasma dynamics are practically identical to FLASH for $n_{\rm max}\geq 5 n_{\rm cr}$ cases, with the main difference being the temperature profile close to the solid target.}
    \label{fig:nmaxscan}
\end{figure}

Figure~\ref{fig:nmaxscan} depicts the results of the $n_{\rm max}$ scan ($n_{\rm max}/n_{\rm cr}=\{2,5,10,20\}$), comparing electron density and temperature profiles with those from FLASH simulations at 0.8 ns. The underdense plasma density evolves similarly in all cases. However, electron temperature matches only when $n_{\rm max}\geq 5n_{\rm cr}$. Lower target densities also result in increased target speeds in the $-z$ direction, which leads to unphysical particle reflection from the simulation boundary at $z=-5 d_i$. Higher values of $n_{\rm max}$ also lead to sharper electron temperature slopes close to the solid target, tending towards the plasma profiles obtained by FLASH (red line).

\begin{figure}
    \centering
    \includegraphics[width=\linewidth]{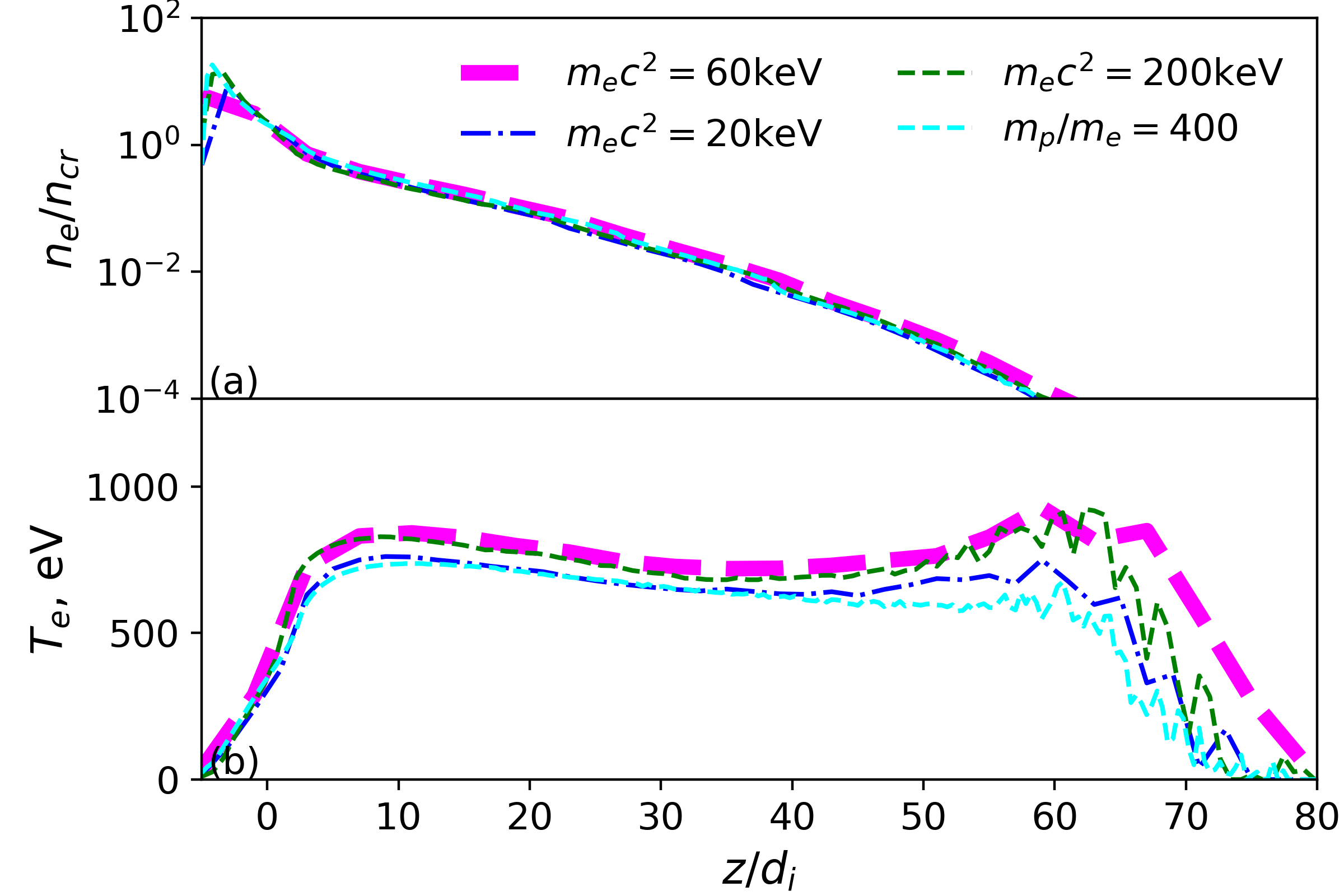}
    \caption{$m_ec_\ast^2$ and $m_p/m_e$ convergence tests showing (a) electron density and (b) electron temperature profiles. The agreement among the resulting plasma profiles justifies using reduced parameters for our PIC simulation.}
    \label{fig:mec2scan}
\end{figure}

Figure~\ref{fig:mec2scan} compares simulations that use different reduced speed of light parameter and ion to electron mass ratio ($m_ec_\ast^2 = \{20,60,200\} \rm keV$; $m_p/m_e = \{100,400\}$). The simulation from the body of this paper used $m_ec_\ast^2 = 60\, \rm keV$ with $m_p/m_e=100$ and additional simulations performed here use $m_ec_\ast^2 = 20$ and 200 keV and an increased mass ratio of $m_p/m_e=400$. All simulations are compared at $0.5$ ns. These simulations demonstrate good convergence, which justifies our use of reduced speed of light and mass ratio parameters.

\section{Electron-ion thermalization test in 1D PIC}\label{sec:eitest}

\begin{figure}
    \centering
    \includegraphics[width=\linewidth]{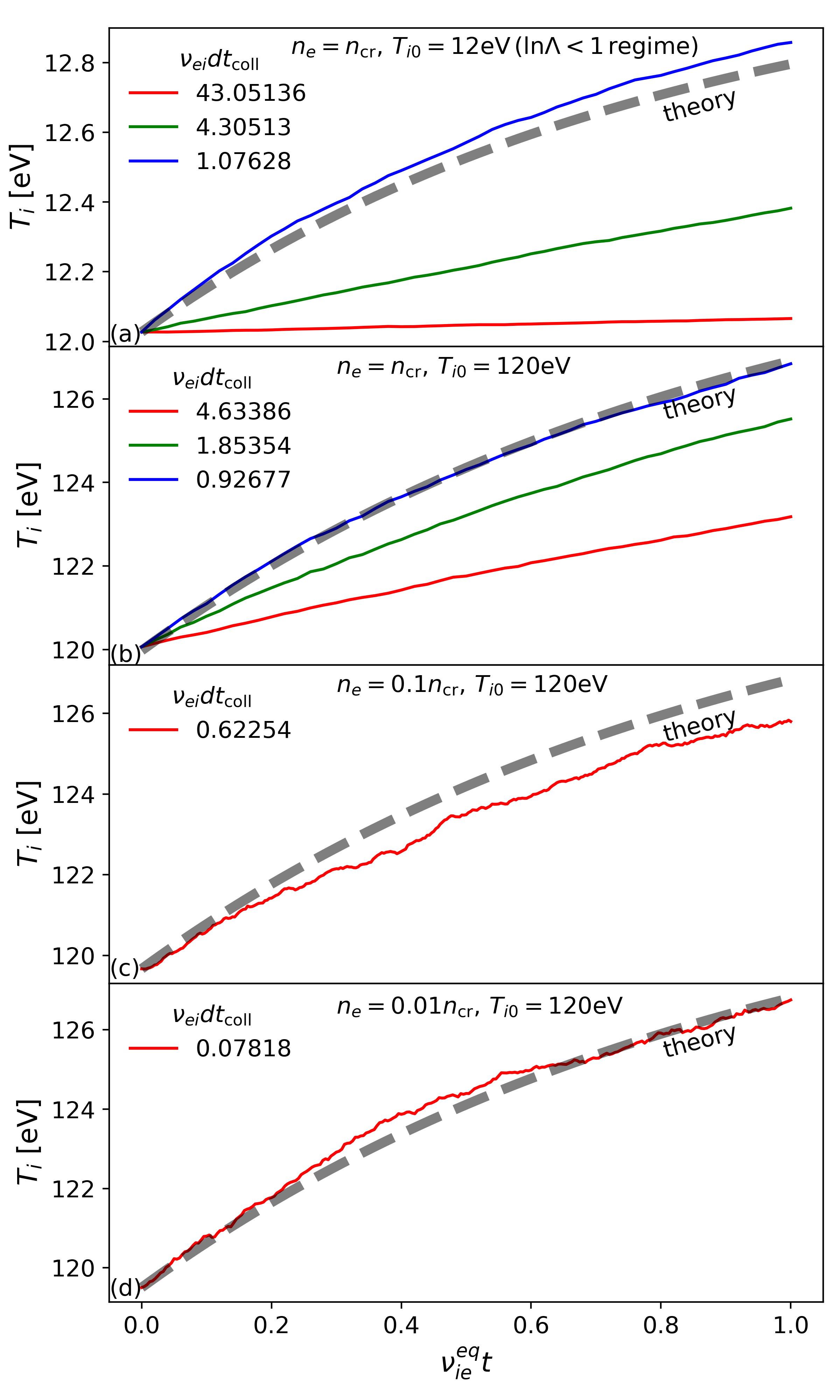}
    \caption{Electron-ion collisional equilibration tests at different plasma conditions and numerical parameters: (a) high density and low temperature: $n_e=n_{\rm cr},\,T_i = 12\rm eV$, $\nu_{ei}dt_{\rm coll} \approx  {43.051,\, 4.305,\, 1.076}$; (b) high density and high temperature: $n_e=n_{\rm cr},\,T_i = 120\rm\, eV$, $\nu_{ei}dt_{\rm coll} \approx  {4.633,\, 1.854,\, 0.926}$; (c) mid density and high temperature: $n_e=0.1n_{\rm cr},\,T_i = 120\rm eV$, $\nu_{ei}dt_{\rm coll} \approx 0.623$; (d) low density and high temperature: $n_e=0.01n_{\rm cr},\,T_i = 120\rm\, eV$, $\nu_{ei}dt_{\rm coll} \approx 0.078$.}
    \label{fig:eieq}
\end{figure}

To verify the validity of the collision module in PSC, we conduct tests of the electron-ion collisional thermalization. We consider a small 1D simulation box ($2d_i$ long) with periodic boundary conditions and use the same resolution parameters as in the main simulation (5 grid cells/$d_e$, $N_{\rm PPC} = 10^5$ particles per cell per species at $n_{e}=n_{\rm cr}$, see Section~\ref{sec:setup}) and the same reduced parameters ($m_p/m_e=100$, $m_ec^2_\ast = 60\, \rm keV$). We then initialize a spatially uniform plasma of Maxwellian electrons and ions. The density and ion temperature is varied across different simulations. The densities are set to $n_e = Z n_i = \{10^{-2},\,10^{-1},1 \}\ n_{\rm cr}$ and the ion temperatures are $T_i=\{12, 120\}\, \rm eV$. The electron temperature is $T_e=1.1 T_i$, with ionization state $Z=13$ and $m_{\rm i}/m_p = 26.7$ corresponding to an aluminum plasma. In this setup, electron-ion collisions should be the dominant process that equilibrates electron and ion populations according to the second term on the RHS of Eq.~\ref{eqn:eion}. We thus compare the evolution of $T_i$ from the PSC simulations with the analytical solution:

\begin{equation}
    T_i(t) = T_{i0} + \frac{Z(T_{e0}-T_{i0})}{Z+1}\left(1-\exp{\left[-\frac{1+Z}{Z} \nu_{\rm ie}^{\rm eq} t\right]} \right).
\end{equation}
\noindent Here, $\nu_{ie}^{\rm eq} = Z(\tau_{ei}^{\mathcal{E}})^{-1}$ and  $\tau_{ei}^{\mathcal{E}}$ are given by Eq.~\ref{eqn:tauei}. We also vary the numerical parameter controlling the frequency of calls to the collision routine in regular PIC timesteps (denoted as $\nu_{ei}dt_{\rm coll}$, with $\nu_{ei}$ being the Braginskii electron-ion collision frequency) to demonstrate convergence and highlight the issues resolving the theoretical equilibration rate at large densities and small temperatures.

Figure~\ref{fig:eieq} presents a set of these tests for variable electron densities, ion temperatures, and frequencies of collision routine calls. The parameters are chosen to probe plasma conditions relevant to the primary PSC simulation discussed in the main text, see Figs.~\ref{fig:flash_psc_evol}a,b. Fig.~\ref{fig:eieq}a uses a high density of $n_e=n_{\rm cr}$, a low temperature of $T_i = 12\rm\, eV$, and varies $\nu_{ei}dt_{\rm coll}$ between 43 (red), 4.3 (green), and 1.1 (blue). These parameters probe the $\ln \Lambda<1$ regime. In this case, the PSC equilibration rate underestimates the theoretical rate unless the collisional routine is called multiple times per PIC timestep (blue line in Fig.~\ref{fig:eieq}a). Figs.~\ref{fig:eieq}b-d all use an increased ion temperature of $T_i = 120\rm eV$, but sequentially decrease the electron density from $n_e=n_{\rm cr}$ in Fig.~\ref{fig:eieq}b to $n_e=0.01n_{\rm cr}$ in Fig.~\ref{fig:eieq}d. Fig.~\ref{fig:eieq}b also shows an underestimation of the theoretical equilibration rate if $\nu_{\rm ei}dt_{\rm coll} > 1$. This may be explained on the basis of the collision module validity criterion formulated in Ref.~\cite{Germaschewski2016}, $\nu_{\rm ei}dt_{\rm coll} \ll 1$. In the regimes with $\nu_{\rm ei}dt_{\rm coll} >1$, the collision frequency is not resolved and the transport coefficients are not accurately reproduced. Reducing the collision timestep to a value satfisfying $\nu_{\rm ei}dt_{\rm coll} <1$ improves the collision module performance, as shown by the blue lines in Figs.~\ref{fig:eieq}a,b. Thus, applying the collision routine every PIC timestep ($dt_{\rm coll} = dt_{\rm PIC}$) would resolve the equilibration rate for the majority of plasma conditions expected in the ablation problem. However, our primary run utilizes $dt_{\rm coll} =10 dt_{\rm PIC}$ to keep the simulation computationally feasible. This underestimates plasma collisionality at $n_e\sim n_{\rm cr}$, but captures collisions correctly at $n_e \sim 0.1 n_{\rm cr}$ or less. The agreement between FLASH and PSC in terms of $T_e/T_i$ across the expanding plasma profiles (see Fig.~\ref{fig:kinetic_effects}b) further affirms the validity of the PSC collision module with $dt_{\rm coll} =10 dt_{\rm PIC}$.

\section{Electron thermal conductivity test in 1D PIC}\label{sec:thermaltest}

\begin{figure}
    \centering
    \includegraphics[width=\linewidth]{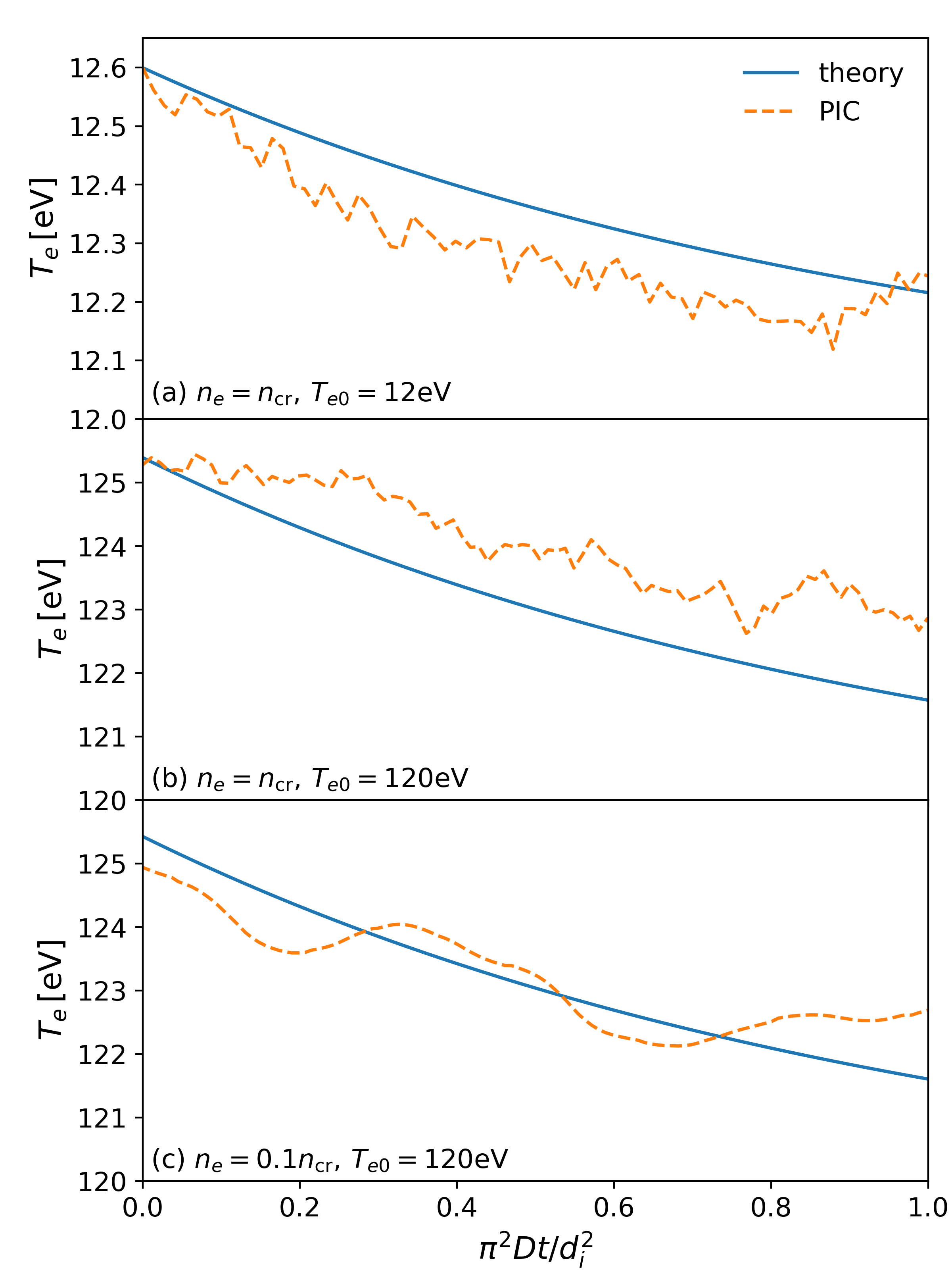}
    \caption{Plasma thermalization tests in 1D PIC. Electron temperature evolution for (a) $n_e=n_{\rm cr}$, $T_{e0}=12\rm eV$, (b) $n_e=n_{\rm cr}$, $T_{e0}=120\rm eV$, (c) $n_e=0.1n_{\rm cr}$, $T_{e0}=120\rm eV$. All cases show good agreement with theory.}
    \label{fig:therm_test}
\end{figure}

Another relevant collision test for the ablation problem is the thermal conduction test. We set up the simulation in the same way as Appendix~\ref{sec:eitest}. We consider a small 1D simulation box ($2d_i$ long) with periodic boundary conditions and use the same resolution parameters as in the main simulation (5 grid cells/$d_e$, $N_{\rm PPC} = 10^5$ particles per cell per species at $n_{e}=n_{\rm cr}$, see Section~\ref{sec:setup}), same reduced units parameters ($m_p/m_e=100$, $m_ec^2_\ast = 60 \rm keV$), and initialize uniform plasma slabs of Maxwellian electrons and ions with densities $n_e = Z n_i =\{0.1, 1\} n_{\rm cr}$ and temperatures $T_{e0}=T_{i0}=\{12,120\} \, \rm eV$, with $Z=13$ and $m_{\rm Al}/m_p = 26.7$ (aluminum plasma). In addition to the bulk temperature, we add a small electron temperature perturbation, which has a sine-wave dependence: $T_e = T_{e0}+ \delta T_{e0} \sin{(\pi z/d_{i})}$, with $\delta T_{e0}/T_{e0} = 0.05$. We expect thermal diffusion, governed by the second to last term on the RHS of Eq.~\ref{eqn:eele} and Eq.~\ref{eqn:heatflux}, to thermalize the plasma according to the following equation:

\begin{equation}
    \rho \frac{\partial}{\partial t}e_e = \nabla \cdot K_e \nabla T_e.
\end{equation}

\noindent Here, $\rho = m_{e} n_e$ is the mass density of electrons, $e_e = 3k_B T_e/m_e$ is the specific internal energy of electrons, $K_e$ is the thermal diffusivity given by Eq.~\ref{eqn:Kele}. In the limit of small temperature perturbation ($K_e \approx \rm const$) and this transforms into a simple 1D diffusion equation:

\begin{eqnarray}
    \frac{\partial}{\partial t}T_e = D \frac{\partial^2}{\partial z^2} T_e, \\
    D \equiv \frac{2 K_e}{3k_Bn_e}.
\end{eqnarray}

\noindent 
For a sine-wave initial condition, $\delta T_e (t=0,z) = \delta T_{e0}\sin{(\pi z/d_i)}$, the system should evolve according to the following solution:

\begin{equation}
    \delta T_e(t,z) = \delta T_{e0}\sin{(\pi z/d_i)} \exp{\left( - \frac{\pi^2 D t}{d_i^2} \right)},
    \label{eqn:thermal_solution}
\end{equation}

\noindent which may be directly compared to PIC simulations.

Figure~\ref{fig:therm_test} summarizes the results of the thermalization tests. Here, we show the evolution of electron temperature at the peak of the sine wave perturbation (shown in dashed orange line), at $z=0.5 d_i$, along with the theoretical estimate of the temporal evolution of electron temperature given by Eq.~\ref{eqn:thermal_solution} (shown in blue line). Good agreement is found for all considered plasma parameters. It should be noted that the thermal transport at higher temperatures or/and lower densities ($n_e \leq 10^{-2}n_{\rm cr}$, $T_e \geq 10^2 \rm \, eV$) reaches a regime that violates the assumptions of classical thermal transport. As shown in Figure~\ref{fig:kinetic_effects}d, $\lambda_{\rm e,mfp}/L_T \sim 1$ at plasma conditions close to $n_e=0.01 n_{\rm cr}$, $T_{e0}=120 \rm eV$. Thermal transport in the non-local regime ($\lambda_{\rm e,mfp}/|L_T|>10^{-2}$) in PSC simulations will be addressed in future work.

\end{document}